\documentclass{cmslatex}
\usepackage[paperwidth=7in, paperheight=10in, margin=.875in]{geometry}
 \usepackage[backref,colorlinks,linkcolor=red,anchorcolor=green,citecolor=blue]{hyperref}
\usepackage{amsfonts,amssymb}
\usepackage{amsmath,bm}
\usepackage{graphicx}
\usepackage{cite}
\usepackage{enumerate}
\renewcommand{\theequation}{\arabic{section}.\arabic{equation}}

\usepackage{mathrsfs}
   \allowdisplaybreaks

\usepackage[utf8x]{inputenc}
\usepackage{esint}
\usepackage[english,french,german,italian,russian]{babel}
\usepackage{appendix}
\def\Fint{\slashint\hspace{-0.5em}\slashint\hspace{-0.5em}\slashint}

\DeclareMathOperator{\Span}{span}
\DeclareMathOperator{\dist}{dist}
\DeclareMathOperator{\ind}{ind}
\DeclareMathOperator{\Tr}{Tr}
\DeclareMathOperator{\Cl}{Cl}
\DeclareMathOperator{\ran}{ran}

\DeclareMathOperator{\Div}{div}
\DeclareMathOperator{\curl}{curl}

\DeclareMathOperator{\D}{d}
\DeclareMathOperator{\I}{Im}

\def\eor{\hfill$ \square$}

\usepackage{enumitem}[2011/09/28]
\setenumerate{align=left, leftmargin=0pt,labelsep=.5em, labelindent=0\parindent,listparindent=\parindent,itemindent=*}

\def\Fint{-\!\!\!-\hspace{-1.65em}\iiint}
\def\ffint{-\!\!\!\!-\hspace{-1.35em}\iiint}

\date{\today}
\usepackage{colonequals}

\usepackage{tikz}
\tikzset{>=stealth}
\usepackage{pgfplots}
\usepackage{color}
\usepackage[normalem]{ulem}
\usepackage[all]{xypic}
\begin{document}

\selectlanguage{english}
\title{Spectral structure of electromagnetic scattering on arbitrarily shaped dielectrics}
\author{Yajun Zhou\thanks{Program in Applied and Computational Mathematics, Princeton University, Princeton, NJ 08544, USA; Academy of Advanced Interdisciplinary Studies (AAIS), Peking University, Beijing 100871, P. R. China, (yajunz@math.princeton.edu, yajun.zhou.1982@pku.edu.cn). } }


\maketitle
\begin{abstract}
Spectral analysis is performed on the Born equation, a strongly singular  integral equation
modeling the interactions between electromagnetic waves and arbitrarily shaped dielectric scatterers. Compact and Hilbert--Schmidt operator polynomials are constructed from the Green operator of electromagnetic scattering on  scatterers with smooth boundaries. As a consequence, it is  shown that the strongly singular Born equation has a discrete spectrum, and that the spectral series $ \sum_\lambda|\lambda|^2|1+2\lambda|^4$ is convergent, counting multiplicities of the eigenvalues $ \lambda$. This reveals a shape-independent optical resonance mode corresponding to a critical dielectric permittivity $ \epsilon_r=-1$.\end{abstract}
\section{Introduction}\subsection{Born equation for electromagnetic scattering on dielectrics}
Consider the dielectric scattering problem in classical electrodynamics, where we shine a monochromatic incident beam (represented by a time-harmonic electric field)  $\bm E_{\mathrm{inc}} (\bm r,t)=\bm E_{\mathrm{inc}} (\bm r)e^{i\omega t} $    onto a homogeneous dielectric occupying a bounded open set $V\Subset \mathbb R^3 $. The dielectric volume $V$ may or may not be connected. For simplicity, we usually assume that the dielectric in question is not hollow, so that  the exterior volume $ \mathbb R^3\smallsetminus(V\cup\partial V)$ is connected.

The dielectric response [\textit{i.e.}~the total electric field $ \bm E(\bm r),\bm r\in V$ inside the dielectric volume] obeys Born's integro-differential equation (see \cite[Kap.~VII]{BornOptik} or \cite[\S13.6.1]{BornWolf}):
\begin{align}\bm E(\bm r)=\bm E_{\mathrm{inc}} (\bm r)+\chi\nabla\times \nabla\times\iiint_V\frac{\bm E(\bm r') e^{-ik|\bm r-\bm r' |}}{4\pi|\bm r-\bm r' |}\D^3 \bm r'-\chi\bm E(\bm r),\quad\bm r\in V.\label{eq:Born}\end{align}
In the \textit{Born equation} \eqref{eq:Born}, the symbol  $\chi=\epsilon_r-1\in\mathbb C$  represents the  susceptibility of the dielectric\footnote{We use the term ``dielectric'' to cover all homogeneous materials with vanishing magnetic susceptibilities. This term incorporates transparent insulators (``dielectrics'' in the narrow sense) with $\chi>-1$, plasmonic materials with $ \chi\leq -1$, and dissipative media with $ \I\chi<0$. The last category also includes  metallic conductors with $ \chi/i<0$.} scatterer relative to the host medium that fills the exterior volume  $ \mathbb R^3\smallsetminus(V\cup\partial V)$, and $2\pi/k$ denotes the wavelength of the incident beam. The three terms on the right-hand side of \eqref{eq:Born} are attributed to, respectively,  the incident wave, the dipole irradiation, and the depolarization process.

In theory,
one can show (see \cite[\S2.6]{Potthast2001} or \cite[\S2]{Kirsch2007}), with mathematical rigor, that the Born equation \eqref{eq:Born} is equivalent to the Maxwell
equations in classical electrodynamics, equipped with   scattering boundary conditions.
In practice, solutions to the Born equation \eqref{eq:Born} enable us to explore the universe surrounding us \cite{LummeBowell,Giese,LeinertGruen} and inspect the
cells within us \cite{ParkRBC,El-Sayed}.

 The Born equation \eqref{eq:Born}, as a linear integro-differential equation, will  be often   abbreviated symbolically as    $ {\hat{\mathscr B}\bm E=(\hat I-\chi\hat{\mathscr G})\bm E=\bm E_\mathrm{inc}}  $ hereafter, with $ \hat {\mathscr B}$ being the Born operator and $ \hat{\mathscr G}$ the Green operator. For a given  combination of the scatterer geometry and beam wavelength, there are  certain special values of the dielectric permittivity  $ \epsilon_r=1+\chi$ that cause \textit{optical resonance}, thus   making the corresponding electromagnetic scattering problem $ (\hat I-\chi\hat{\mathscr G})\bm E=\bm E_\mathrm{inc}$  ill-posed (\textit{i.e.}~the corresponding Born operator $ \hat {\mathscr B}=\hat I-\chi\hat{\mathscr G}$ fails to have a bounded inverse). Characterization of the optical resonance spectrum\footnote{Motivated by the experimental need  \cite{Zheng} to customize permittivities of nanophotonic probes that are excited by
a monochromatic laser,
we are considering  a scattering problem with fixed wavelength $ 2\pi/k\in\mathbb R$ for the incident beam and variable permittivities $ \epsilon_r=1+\chi\in\mathbb C$ inside the dielectric volume. Motivated by numerical solutions to the Maxwell equations, one may also consider scattering problems with a complex variable $ k\in\mathbb C$ \cite{Kirsch2007,Costabel2010,Chesnel2012}, while  fixing the spatial distributions of the dielectric permittivity $ \epsilon_r(\bm r)$ and the magnetic permeability $ \mu_r(\bm r)$.
Thus, our conclusions about the $ \chi$-spectrum (such as discrete eigenvalues in Theorem \ref{thm:PC} below) and previous studies of the $k$-spectrum (such as discrete eigenvalues in \cite[Theorem 3.7]{Kirsch2007})
are  mathematically inequivalent and are of independent scientific interests.} is
the major task of our present paper.

To make our spectral analysis mathematically precise, we need to clearly specify the Hilbert space in which we want to solve  the Born equation $ (\hat I-\chi\hat{\mathscr G})\bm E=\bm E_\mathrm{inc}$. A na\"ive choice for such a Hilbert space is the totality of \textit{energetically admissible} electric fields
$L^2(V;\mathbb C^3)$,  consisting of square-integrable $ \mathbb C^3$-valued vector fields $ \bm E(\bm r),\bm r\in V$ defined in the dielectric volume $V$.
Both  $ \hat {\mathscr B}\colon L^2(V;\mathbb C^3)\longrightarrow L^2(V;\mathbb C^3)$ and  $ \hat{\mathscr G}\colon L^2(V;\mathbb C^3)\longrightarrow L^2(V;\mathbb C^3)$ are bounded linear operators.
A more circumspect choice
is a proper subspace of $ L^2(V;\mathbb C^3)$, the totality of \textit{physically admissible} electric fields
$\Phi(V;\mathbb C^3)\colonequals \Cl(C^\infty(V;\mathbb C^3)\cap\ker(\nabla\cdot)\cap L^2(V;\mathbb C^3))$, which  is the $ L^2$-closure for all smooth, divergence-free and square-integrable  $ \mathbb C^3$-valued vector fields  $ \bm E(\bm r),\bm r\in V$. A conventional notation for $\Phi(V;\mathbb C^3)$ is $ H(\Div0 ,V)$ \cite[p.~215]{DautrayLionsVol3}.

The Born equation $ (\hat I-\chi\hat{\mathscr G})\bm E=\bm E_\mathrm{inc}$ on the Hilbert space
$L^2(V;\mathbb C^3)$  is effectively a strongly singular integral equation, whose integral kernel exhibits $ O(|\bm r-\bm r'|^{-3})$ divergence at short distances. Such a strong singularity disqualifies the Green operator  $ \hat{\mathscr G}\colon L^2(V;\mathbb C^3)\longrightarrow L^2(V;\mathbb C^3)$ as a compact operator, which makes the Riesz--Schauder theory  \cite{Richtmyer,Yosida} for the Fredholm integral equations not directly applicable to the Born equation \eqref{eq:Born}.

When we restrict to the physically meaningful Hilbert subspace  $ \Phi(V;\mathbb C^3)$, we will prove in this work  that   $ \hat{\mathscr  G}(\hat I+2\hat{\mathscr G)}\colon \Phi(V;\mathbb C^3)\longrightarrow\Phi(V;\mathbb C^3)$ is a compact operator and that   $ \hat{\mathscr  G}(\hat I+2\hat{\mathscr G})^{2}\colon \Phi(V;\mathbb C^3) \longrightarrow\Phi(V;\mathbb C^3)$ is a Hilbert--Schmidt operator, so long as the dielectric boundary $ \partial V$ is smooth. In less abstract terms, one can show that    $ \hat{\mathscr  G}(\hat I+2\hat{\mathscr G)}\colon \Phi(V;\mathbb C^3)\longrightarrow\Phi(V;\mathbb C^3)$ is equivalent to a weakly singular integral kernel with $ O(|\bm r-\bm r'|^{-2})$ asymptotic behavior, and that    $ \hat{\mathscr G }(\hat I+2\hat{\mathscr G})^2\colon \Phi(V;\mathbb C^3)\longrightarrow\Phi(V;\mathbb C^3)$ is comparable to a square-integrable integral kernel that diverges as  $ O(|\bm r-\bm r'|^{-1})$ at short distances.
\subsection{Statement of results and plan of proof}
Following standard practices   \cite{Richtmyer,Yosida}  in functional analysis,  we write   $ \sigma^\Phi(\hat{\mathscr G})=\sigma^\Phi_p(\hat{\mathscr G})\cup \sigma^\Phi_c(\hat{\mathscr G})\cup\sigma^\Phi_r(\hat{\mathscr G})$ for the spectrum [\textit{i.e.}~complement of the resolvent set $\rho^\Phi(\hat{\mathscr G}) $] of    $ \hat{\mathscr  G}\colon \Phi(V;\mathbb C^3)\longrightarrow\Phi(V;\mathbb C^3)$, which decomposes into the point spectrum $ \sigma^\Phi_p(\hat{\mathscr G})$ (\textit{i.e.}~the set of eigenvalues), the continuous spectrum $ \sigma^\Phi_c(\hat{\mathscr G})$, and the residual spectrum $ \sigma^\Phi_r(\hat{\mathscr G})$.  These notations allow us to formally state our main results in the next two theorems.\begin{theorem}[Compact polynomial and optical resonance]\label{thm:PC}
Suppose that the dielectric volume is a bounded open set $ V\Subset \mathbb R^3$ with smooth boundary $ \partial V$, then
the following properties hold true.

\begin{description}
\item[Polynomial compactness]
The non-compact Green operator  $\hat{\mathscr G }\colon \Phi(V;\mathbb C^3)\longrightarrow\Phi(V;\mathbb C^3)$ is polynomially compact, with minimal polynomial $ \hat{\mathscr G }(1+2\hat{\mathscr G })/2$.
\item[Decomposition of spectrum] The residual spectrum is empty $\sigma^\Phi_r(\hat{\mathscr G})=\varnothing $, and the physical spectrum  $ \sigma^\Phi(\hat{\mathscr G})$ is the union of the continuous spectrum $ \sigma^\Phi_c(\hat{\mathscr G})=\{0,-1/2\}$ with the point spectrum $ \sigma^\Phi_p(\hat{\mathscr G})$ that contains countably many eigenvalues. \item[Characterization of eigenvalues] Each eigenvalue     $ \lambda\in\sigma^\Phi_{p}(\hat{\mathscr G})$
 is associated with a finite-dimensional eigen\-space, and $ \I\lambda\leq0$. Furthermore, if the exterior volume $ \mathbb R^3\smallsetminus(V\cup\partial V)$ is connected, then each eigenvalue    $ \lambda\in\sigma^\Phi_{p}(\hat{\mathscr G})$
has a strictly negative imaginary part $ \I\lambda<0$.
\item[Continuous spectrum and universal resonance]
The continuous spectrum  $ \sigma^\Phi_c(\hat{\mathscr G})=\{0,-1/2\}$ forms the only possible accumulation points of eigenvalues.
The shape-independent singularity $-1/2\in\sigma^\Phi_c(\hat{\mathscr G})  $ corresponds to a universal optical resonance\footnote{The unbounded inverse of $ \hat I+2\hat{\mathscr G}\colon \Phi(V;\mathbb C^3)\longrightarrow\Phi(V;\mathbb C^3) $ points to a field enhancement (for  $\epsilon_r=-1$)  $\sup_{\bm E_{\mathrm {inc}}\neq\mathbf 0}\iiint_V|\bm E(\bm r)|^{2}\D^3\bm r/\iiint_V|\bm E_{\mathrm{inc}}(\bm r)|^{2}\D^3\bm r=+\infty\ $ that is independent of scatterer geometry. This contrasts with other geometry-dependent enhancement effects in the electrostatic approximation \cite{Geom2D,3DconvEPS} to electromagnetic scattering.} at susceptibility $ \chi=-2$, i.e.~relative permittivity $\epsilon_r=-1$.
\end{description}

 \end{theorem}

\begin{theorem}[Hilbert--Schmidt polynomial and spectral series]\label{thm:HS}For a bounded dielectric volume $ V\Subset\mathbb R^3$ with smooth boundary $ \partial V$, the  operator  $\hat{\mathscr G }(\hat I+2\hat{\mathscr G})^2\colon \Phi(V;\mathbb C^3)\longrightarrow\Phi(V;\mathbb C^3)$ is of Hilbert--Schmidt type, and we have a convergent spectral series\begin{align} \sum_{\lambda\in\sigma^\Phi(\hat{\mathscr G})}|\lambda|^2|1+2\lambda|^4<+\infty,\end{align} where the sum respects multiplicities in eigenvalues.\end{theorem}

A diligent reader may detect subtle differences between the statements in our last two theorems and reports from previous  studies of electromagnetic scattering \cite{Rahola,Budko,Budko2007,HsiaoKleinman,CostabelDarrigrandSakly2012}, in at least two different ways. \begin{itemize}
\item

Rahola \cite{Rahola}  had proposed a hypothesis about the existence of a continuum of optical resonance modes across  the plasmon range $ \epsilon_r<0$, contrary to our discrete spectrum.\footnote{The lack of resonance modes in the plasmon range $ \epsilon_r<0$ in our spectrum should not be misconstrued as counterevidence for  \textit{plasmonic resonance}, an experimentally observed physical phenomenon. A proper mathematical analysis of the plasmonic resonances in nanoparticles requires the full Maxwell equations \cite{Ammari2016}, with tunable electric permittivities $ \epsilon_r$ and magnetic permeabilities $ \mu_r$, which is beyond the scope of the current research.} In addition to supporting Rahola's hypothesis, Budko and Samokhin \cite{Budko,Budko2007} had speculated about a resonance mode at $ \epsilon_r=0$, which is absent from our spectral analysis in Theorem \ref{thm:PC}. The main reason for the broader span of  their spectra is their choice of a larger Hilbert space [namely $L^2(V;\mathbb C^3)$ instead of $ \Phi(V;\mathbb C^3)$, cf.~Theorem \ref{thm:sp_L2} below] that does not necessarily honor the physical constraint of transversality  $ \nabla\cdot\bm E(\bm r)=0,\bm r\in V$.\footnote{One observes that the Born equation  \eqref{eq:Born} brings us $ \nabla\cdot[(1+\chi)\bm E(\bm r)]=\nabla\cdot\bm E_{\mathrm{inc}} (\bm r)=0,\bm r\in V$, which immediately implies transversality   $ \nabla\cdot\bm E(\bm r)=0,\bm r\in V$ for $\chi\neq-1$. We still impose the transversality constraint while investigating the spectral r\^ole of $\chi=-1$, because realistic materials (including transparent media like water and glass) absorb light and dissipate electromagnetic energy through a non-vanishing $ \I\chi$ (no matter how small it is), making  $\chi=-1 $ practically inaccessible---when we are inquiring about the spectral status of a real-valued $\chi_{r}$, we are effectively asking whether the electric enhancement ratio $ \sup_{\bm E_{\mathrm {inc}}\neq\mathbf 0}\iiint_V|(\hat I-\chi\hat{\mathscr G})^{-1}\bm E_{\mathrm{inc}}(\bm r)|^{2}\D^3\bm r/\iiint_V|\bm E_{\mathrm{inc}}(\bm r)|^{2}\D^3\bm r$ goes unbounded as $ \chi\to\chi_r-i0^+$. During the limit procedure  $ \chi\to\chi_r-i0^+$, one can always divide by a non-vanishing $1+\chi$ and justify the transversality condition for $\Phi(V;\mathbb C^3)$. Therefore, it is legitimate to use the same Hilbert space $\Phi(V;\mathbb C^3) $ for physically admissible electric fields, no matter $\chi=-1$ or not. } The non-physical nature of the conjectural arguments by Rahola  and Budko--Samokhin have been noted by  the present author \cite{ZhouThesis}, Costabel--Darrigrand--Sakly \cite{CostabelDarrigrandSakly2012} and Zouros--Budko \cite{ZourosBudko2012}.\item Using a different  function space $H(\curl, V) $ [requiring square-integrability of both $ \bm E(\bm r)$ and $\nabla\times \bm E(\bm r) $],\footnote{It is worth pointing out that for numerical analysis of the Maxwell equations, the Hilbert space $H(\curl, V) $ is by far a more popular choice than $ H(\Div 0,V)$ \cite{Monk2003}.} Hsiao--Kleinman \cite[Sect.~VI]{HsiaoKleinman}  and Costabel--Darrigrand--Sakly \cite[Corollary 3.3]{CostabelDarrigrandSakly2012}  arrived at similar conclusions about the presence of shape-independent resonance at  $ \chi=-2$ for 3-dimensional electromagnetic scattering, while working with a surface integral formulation. For aesthetic reasons, we are not imposing an $L^2$-constraint on the time-harmonic magnetic field [which is proportional to  $ \nabla\times \bm E(\bm r) $] in our present work, since that points to a stronger norm of the electric field than our choice [where the norm of $ H(\Div 0,V)$ inherits that of $ L^2(V;\mathbb C^3)$]. In our analysis, we find that the transversality constraint $\nabla\cdot \bm E(\bm r)=0 $ is a minimal setting to guarantee a discrete spectrum $ \sigma^\Phi(\hat{\mathscr G})$ and  a shape-independent resonance $-1/2\in \sigma_c^\Phi(\hat{\mathscr G}) $. \end{itemize}

In contrast to the Born equation \eqref{eq:Born}, the acoustic scattering equation in three-dimensional space for a scalar field $ u(\bm r),\bm r\in V$ reads\begin{align}((\hat I-\chi k^2\hat {\mathscr C})u)(\bm r)\colonequals u(\bm r)-\chi k^2\iiint_V\frac{u(\bm r')e^{-ik|\bm r-\bm r'|}}{4\pi|\bm r-\bm r'|}\D^3\bm r'=u_\mathrm{inc}(\bm r), \quad u_\mathrm{inc}\in L^2(V;\mathbb C).\label{eq:Born'}\tag{\ref{eq:Born}$'$}\end{align} This involves a  perturbation of the identity operator  $\hat I $ by a weakly singular operator \linebreak $ -\chi k^2\hat{\mathscr C}\colon L^2(V;\mathbb C)\longrightarrow L^2(V;\mathbb C)$ with $ O(|\bm r-\bm r'|^{-1})$ divergence at short distances, and thus has significantly simplified spectral properties as compared to the Born equation for electromagnetic scattering.
In particular, via the  Riesz--Schauder theory and the square integrability of the weakly singular kernel  with $ O(|\bm r-\bm r'|^{-1})$ behavior, we can readily confirm the following analogs of Theorems \ref{thm:PC} and \ref{thm:HS}:\begin{enumerate}[leftmargin=*,  label=(\arabic*$'$),
widest=2, align=left]
\item The {spectrum} $ \sigma(\hat{\mathscr C})$ of the  compact operator $ \hat{\mathscr C}\colon L^2(V;\mathbb C)\longrightarrow L^2(V;\mathbb C)$ consists of countably many eigenvalues in the point spectrum $ \sigma_p(\hat{\mathscr C})$, and $ \{0\}\subset\sigma_c(\hat{\mathscr C})$ being the only possible accumulation point.\item For the Hilbert--Schmidt operator  $ \hat{\mathscr C}$,  the spectral series $ \sum_{\lambda\in\sigma(\hat {\mathscr C})}|\lambda|^2$ is convergent.  \end{enumerate}

After opening with a discussion on the spectrum $ \sigma(\hat{\mathscr G})=\sigma_p(\hat{\mathscr G})\cup \sigma_c(\hat{\mathscr G})\cup\sigma_r(\hat{\mathscr G})=\mathbb C\smallsetminus\rho(\hat{\mathscr G})$ of $ \hat{\mathscr G}\colon L^2(V;\mathbb C^3)\longrightarrow  L^2(V;\mathbb C^3)$ in \S\ref{sec:L2_spec},  we will verify Theorem~\ref{thm:PC}  in its entirety in \S\S\ref{subsec:uniq_eps0}--\ref{subsec:cpt_poly}.
As by-products, we will also prove some  results for $ \sigma(\hat{\mathscr G})$ (as stated in the next theorem) in parallel to Theorem~\ref{thm:PC}.

\begin{theorem}[Generalized optical theorem and its consequences]\label{thm:sp_L2}When the dielectric volume $ V$ is a bounded open set, the Born operator $\hat{\mathscr B}=\hat I-\chi\hat {\mathscr G}\colon L^2(V;\mathbb C^3)\longrightarrow L^2(V;\mathbb C^3) $ satisfies the following generalization of   the standard optical theorem \cite[\S10.11]{Jackson:EM}: \begin{align}&
\frac{|\chi|^2 k^5}{16\pi^2 } \varoiint_{|\bm n|=1}\left|\bm n\times\iiint_V \bm E(\bm r' ) e^{ik\bm n \cdot \bm r' }  \D^3\bm r' \right|^2\D\Omega\notag\\={}&\I\left[\chi k^2\iiint_V|\bm E(\bm r )|^{2}\D^3\bm r \right]-\I\left[\chi k^2\iiint_V(\hat{\mathscr  B}\bm E)^*(\bm r)\cdot\bm E(\bm r )\D^3\bm r\right],\label{eq:GOT1}
\end{align}where $ \D\Omega=\sin\theta\D\theta\D\phi$ stands for the surface element on the unit sphere $ |\bm n|=1$, and an asterisk denotes complex conjugation. From this functional relation one can deduce the following properties when the dielectric volume is a bounded open set $ V\Subset \mathbb R^3$ with smooth boundary $ \partial V$.\footnote{To facilitate comparisons with Theorem~\ref{thm:PC}, we state our results here for bounded and open dielectric volumes $V$ with smooth boundaries $ \partial V$. Such geometric requirements can be relaxed (say, down to ``bounded and open volumes  $V$ whose boundaries  $ \partial V$ occupy zero volumes''), as we progress to the actual proof of Theorem~\ref{thm:sp_L2} in \S\ref{sec:L2_spec}. Thus, the phrase ``arbitrarily shaped'' in the title of this work is meant to cover both the smooth dielectrics and the non-smooth ones.} \begin{description}
\item[Characterization of eigenvalues]
  If  $ \lambda\in\sigma_p(\hat{\mathscr G})$
is an eigenvalue of $ \hat {\mathscr G}\colon L^2(V;\mathbb C^3)\longrightarrow L^2(V;\mathbb C^3)$, then   $ \I\lambda\leq0$. If we know additionally that $ \mathbb R^3\smallsetminus(V\cup\partial V)$ is connected, then each eigenvalue also satisfies
$\lambda\notin(-\infty,-1)\cup(-1,+\infty)$.\item[Absence of residual spectrum] We have $ \sigma_r(\hat{\mathscr G})=\varnothing$.
\item[Non-physical resonance mode at $ \epsilon_r=1+\chi=0$] We have $ 1/\chi=-1\in\sigma_p(\hat{\mathscr G})$ as an eigenvalue of infinite multiplicity $ \dim\ker(\hat I+\hat{\mathscr G})=+\infty$.\item[Non-physical resonance modes for $ \epsilon_r=1+\chi<0$]
 If  $ \mathbb R^3\smallsetminus(V\cup\partial V)$ is connected, then $ (-1,0]\subset\rho(\hat{\mathscr G})\cup\sigma_c(\hat{\mathscr G})$.\end{description}\end{theorem}
Using the surface convolution formulation of the Green operator  $ \hat{\mathscr G}$ and some careful manipulations of multiple surface integrals, we will give an extension of the polynomial compactness in \S\ref{subsec:HS_poly}, thereby proving Theorem \ref{thm:HS}.

The present work only touches upon some qualitative aspects of the spectral structures in electromagnetic scattering. Some conclusions in this article will also become useful when we shift our focus to quantitative  spectral analysis of electromagnetic scattering \cite{QuantEM1Norm,QuantEM2Evolv}.

\section{Energy condition and generalized optical theorem\label{sec:L2_spec}}

In this section, we begin with some preparations in Fourier analysis (\S\ref{subsec:FT}) that enable us to rewrite  the Born equation     $ {\hat{\mathscr B}\bm E=(\hat I-\chi\hat{\mathscr G})\bm E=\bm E_\mathrm{inc}}  $   in the wave-vector space (for the Fourier transform of electric fields). We then address the  uniqueness issue (\S\ref{subsec:uniq_L2}) and the existence issue (\S\ref{subsec:exist_L2})  separately, while discussing the $L^2$-solvability of the electromagnetic scattering problem.
Here, for the uniqueness theorem, we need to check that the kernel space $ \ker(\hat I-\chi\hat{\mathscr G})$ is trivial for a certain value of   $\chi$; for the existence theorem, we need to prove that the range $ \ran(\hat I-\chi\hat{\mathscr G})$ occupies the entire function space of energetically admissible electric fields
for a given  susceptibility $\chi  $. Putting together uniqueness and existence, we can say that  the Born operator  $ \hat{\mathscr B}=\hat I-\chi\hat{\mathscr G}\colon L^2(V;\mathbb C^3)\longrightarrow  L^2(V;\mathbb C^3) $    has a bounded inverse, and the corresponding Born equation is   $L^2$-solvable.

During the rest of this section (\S\S\ref{subsec:FT}--\ref{subsec:exist_L2}), we will focus on  the consequence of the energy constraint:  $ \iiint_V|\bm E(\bm r)|^2\D^3\bm r<+\infty$. The physical requirement of transversality $ \nabla\cdot\bm E=0$ will not be imposed until the next section (\S\ref{sec:div0}). Unlike most results in \S\ref{sec:div0} that will hinge on the smoothness of the dielectric boundary $ \partial V$, the analysis in the current section typically only requires the dielectric volume $V$ to be a bounded open subset of $ \mathbb R^3$. \textit{Caveat lector}:  although certain results  in this section  apply equally well to the Green operator hitting on the physical Hilbert space  $  \hat{\mathscr G}\colon \Phi(V;\mathbb C^3)\longrightarrow \Phi(V;\mathbb C^3)$ (such as bounded operator norms), the lack of the transversality constraint  $ \nabla\cdot\bm E=0$ causes a lot of non-physical artifacts in the spectral analysis of  $  \hat{\mathscr G}\colon L^2(V;\mathbb C^3)\longrightarrow L^2(V;\mathbb C^3)$, so not every  point in this $L^2$-spectrum  corresponds to  physical ill-posedness of the electromagnetic scattering problem.
\subsection{Fourier analysis for the Born equation\label{subsec:FT}}We
start by decomposing the Green operator $ \hat{\mathscr G}$ into the sum of two linear operators (with the understanding that all the partial derivatives are interpreted in the distributional sense hereinafter, unless explicitly specified otherwise):\begin{align}
(\hat\gamma\bm E)(\bm r)\colonequals \nabla\times \nabla\times\iiint_V\frac{{\bm E(\bm r')} (e^{-ik|\bm r-\bm r' |}-1)}{4\pi|\bm r-\bm r' |}\D^3 \bm r'\label{eq:defn_gamma_op}
\end{align}and \begin{align}
((\hat{\mathscr G}-\hat\gamma)\bm E)(\bm r)\colonequals \nabla\times \nabla\times\iiint_V\frac{{\bm E(\bm r')} }{4\pi|\bm r-\bm r' |}\D^3 \bm r'-\bm E(\bm r)=\nabla\left[ \nabla\cdot\iiint_V\frac{\bm E(\bm r')}{4\pi|\bm r-\bm r'|} \D^3\bm r'\right].\label{eq:defn_G_g}
\end{align}

Here, it is easy to rewrite the right-hand side of \eqref{eq:defn_gamma_op} as a convolution of the electric field with a square-integrable kernel [with $ O(|\bm r-\bm r'|^{-1})$ behavior at short distances], so  \eqref{eq:defn_gamma_op} defines a Hilbert--Schmidt operator  $ \hat\gamma\colon L^2(V;\mathbb C^3)\longrightarrow L^2(V;\mathbb C^3)$.

We devote the next proposition to a scrutiny of the operator $ \hat{\mathscr G}-\hat\gamma$.

\begin{proposition}[Fourier analysis of $ \hat{\mathscr G}-\hat\gamma$]\label{prop:Gg_FT}\begin{enumerate}[leftmargin=*,  label=\emph{(\alph*)},ref=(\alph*),
widest=a, align=left]
\item The following Fourier inversion formulae hold
for all $ \bm E\in L^2(V;\mathbb C^3)$:\begin{align}&(\bm u\cdot\nabla)(\bm v\cdot\nabla)\iiint_V\frac{\bm E(\bm r' )}{4\pi|\bm r-\bm r' |}  \D^3 \bm r'\notag\\={}&-\frac1{(2\pi)^3}\Fint_{\mathbb R^3}\frac{(\bm u\cdot\bm q)(\bm v\cdot\bm q)\widetilde {\bm E}(\bm q)}{|\bm q|^2}e^{-i\bm q\cdot \bm r}\D^3\bm q,\quad\forall \bm u,\bm v\in\{\bm e_x,\bm e_y,\bm e_z\},\label{eq:Npot_deriv_FT}\end{align}where   \begin{align}\widetilde{\bm F}(\bm q)\colonequals \Fint_{\mathbb R^3}\bm F(\bm r)e^{i\bm q\cdot \bm r}\D^3\bm r\label{eq:fwFT}\end{align}denotes $ L^2$-Fourier transform,\footnote{For $\bm F\in L^1(\mathbb R^3;\mathbb C^3)\cap L^2(\mathbb R^3;\mathbb C^3)  $, the $ L^2$-Fourier transform \eqref{eq:fwFT} coincides with an absolutely convergent  Lebesgue integral $\widetilde{\bm F}(\bm q)=\iiint_{\mathbb R^3}\bm F(\bm r)e^{i\bm q\cdot \bm r}\D^3\bm r$. For a generic  $\bm F\in L^2(\mathbb R^3;\mathbb C^3)  $, its  $ L^2$-Fourier transform \eqref{eq:fwFT}   still has precise meanings as continuous extensions of the Fourier transform from $ L^1(\mathbb R^3;\mathbb C^3)\cap L^2(\mathbb R^3;\mathbb C^3)$ to $L^2(\mathbb R^3;\mathbb C^3) $. (See \cite[p.~104]{Grafakos} for further technical explanations of the operation ``$ \ffint$''.) In general, when ``$ \ffint$'' appears in an equality, it means that the left- and right-hand sides agree with possible exceptions on a subset  $V_0\subset \mathbb R^3$ of zero volume (\textit{i.e.}~$\iiint_{V_0}\D^3\bm r=0$ or $\iiint_{V_0}\D^3\bm q=0$ depending on context).
} whose inverse is $ \bm F(\bm r)=\frac1{(2\pi)^3}\ffint_{\mathbb R^3}\widetilde{\bm F}(\bm q)e^{-i\bm q\cdot \bm r}\D^3\bm q$.\item The right-hand side of  \eqref{eq:defn_G_g} defines a bounded linear operator $ \hat{\mathscr G}-\hat\gamma\colon L^2(V;\mathbb C^3)\longrightarrow L^2(V;\mathbb C^3) $, satisfying\begin{align}
0\leq-\langle\bm E,(\hat{\mathscr G}-\hat\gamma)\bm E \rangle_V\leq\langle\bm E,\bm E \rangle_V,\label{eq:Hermitian_ineq}
\end{align}where $ \langle \bm F,\bm G\rangle_V\colonequals \iiint_V \bm F^*(\bm r)\cdot\bm G(\bm r)\D^3\bm r$ denotes inner product in the Hilbert space $ L^2(V;\mathbb C^3)$.\end{enumerate}\end{proposition}\begin{proof}
\begin{enumerate}[leftmargin=*,  label={(\alph*)},ref=(\alph*),
widest=a, align=left]
\item First, we point out that    a convolution with the Newtonian potential $(4\pi|\bm r-\bm r'|)^{-1} $  in $\bm r $-space can be rendered as a  multiplication of $|\bm q|^{-2} $ in $\bm q $-space \begin{align}(\hat{\mathscr N}\bm F)(\bm r)={}&\iiint_{\mathbb R^3}\frac{\bm F (\bm r' )}{4\pi|\bm r-\bm r' |}  \D^3 \bm r'=\frac1{(2\pi)^3}\iiint_{\mathbb R^3}\frac{\widetilde{ \bm F} (\bm q)}{|\bm q|^2}e^{-i\bm q\cdot \bm r}\D^3\bm q,\quad \bm r\in\mathbb R^3,\notag\\{}&\text{for }\widetilde{ \bm F}(\bm q)\colonequals \iiint_{\mathbb R^3}\bm F (\bm r)e^{i\bm q\cdot \bm r}\D^3\bm r,  \label{eq:Newton_potential} \end{align}
    so long as  $\bm F\in{\mathscr S }(\mathbb R^3;\mathbb C^3)  $ \cite[p.~117, Lemma 1(a)]{SteinDiff}. Here, the Schwartz space is  defined by
  \begin{align}{\mathscr S }(\mathbb R^3;\mathbb C^3)\colonequals \left\{\bm F\in C^\infty(\mathbb R^3;\mathbb C^3)\left|\sup_{\bm r\in \mathbb R^3}\left\vert \bm r^{\bm \mu}D^{\bm \nu}\bm F(\bm r) \right\vert<+\infty, \forall \bm \mu,\bm \nu\right.\right\}\end{align}
where $ \bm r^{\bm \mu}$  is a monomial  $x^{\mu_x}y^{\mu_y}z^{\mu_z}$ of total degree $\mu_x+\mu_y+\mu_z=|\bm \mu| $,  and $ D^{\bm \nu}$ denotes partial derivatives   with respect to the multi-index $\bm \nu$ of total order $|\bm \nu|$. If $\bm F\in{\mathscr S }(\mathbb R^3;\mathbb C^3) $, then $ \widetilde{\bm F}\in{\mathscr S }(\mathbb R^3;\mathbb C^3)$, and vice versa \cite[Corollary 2.2.15]{Grafakos}.
It is clear from the definition that $C^\infty_0(V;\mathbb C^3)\subset{\mathscr S }(\mathbb R^3;\mathbb C^3) $, where $ C^\infty_0(V;\mathbb C^3)$ is the totality of smooth and compactly supported $ \mathbb C^3$-valued  vector fields defined in $V$.

Using  integration by parts, one can justify the continuity of second-order partial derivatives  $\hat {\mathscr N}\bm F\in C^2(\mathbb R^3;\mathbb C^3) $ for all $\bm F\in{\mathscr S }(\mathbb R^3;\mathbb C^3)$. Meanwhile, $\widetilde {\bm F}\in{\mathscr S }(\mathbb R^3;\mathbb C^3)  $ implies the absolute convergence of the following integrals for all $ \bm u,\bm v\in\{\bm e_x,\bm e_y,\bm e_z\}$: \begin{align}\iiint_{\mathbb R^3}\frac{|(\bm u\cdot\bm q)\widetilde{ \bm F} (\bm q)|}{|\bm q|^2}\D^3\bm q<+\infty,\; \iiint_{\mathbb R^3}\frac{|(\bm u\cdot\bm q)(\bm v\cdot\bm q)\widetilde{ \bm F} (\bm q)|}{|\bm q|^2}\D^3\bm q<+\infty, \end{align} and hence the continuity (with respect to $\bm r\in \mathbb R^3 $) of the corresponding Fourier integrals
\begin{align}\begin{split}\iiint_{\mathbb R^3}\frac{(\bm u\cdot\bm q)\widetilde{ \bm F} (\bm q)}{|\bm q|^2}e^{-i\bm q\cdot \bm r}\D^3\bm q\in{}& C^0(\mathbb R^3;\mathbb C^3),\\ \iiint_{\mathbb R^3}\frac{(\bm u\cdot\bm q)(\bm v\cdot\bm q)\widetilde{ \bm F} (\bm q)}{|\bm q|^2}e^{-i\bm q\cdot \bm r}\D^3\bm q\in{}& C^0(\mathbb R^3;\mathbb C^3)\end{split}\end{align}for all $ \bm u,\bm v\in\{\bm e_x,\bm e_y,\bm e_z\}$. All this information combined allows us to take derivatives under the integral sign, according to  the Fubini theorem and the Newton--Leibniz formula, thus leading to the following result
\begin{align}\label{eq:FTinvN}&(\hat{\mathscr D}_{\bm u,\bm v}\bm F)(\bm r)\colonequals (\bm u\cdot\nabla)(\bm v\cdot\nabla)\iiint_{\mathbb R^3}\frac{\bm F(\bm r' )}{4\pi|\bm r-\bm r' |}  \D^3 \bm r\notag\\={}&-\frac1{(2\pi)^3}\iiint_{\mathbb R^3}\frac{(\bm u\cdot\bm q)(\bm v\cdot\bm q)\widetilde {\bm F}(\bm q)}{|\bm q|^2}e^{-i\bm q\cdot \bm r}\D^3\bm q\end{align} for all $\bm F\in C^\infty_0(V;\mathbb C^3)\subset{\mathscr S }(\mathbb R^3;\mathbb C^3)  $ and $ \bm u,\bm v\in\{\bm e_x,\bm e_y,\bm e_z\}$.

We recall that $C^\infty_0(V;\mathbb C^3) $ is dense in $L^2(V;\mathbb C^3) $ \cite[Lemma 2.19]{Lieb}.   For a generic square-integrable vector field $ \bm E\in  L^2(V;\mathbb C^3)$,  let $\{\bm E_s\in C^\infty_0(V;\mathbb C^3) |s=1,2,\dots \}$ be a sequence that converges to $\bm E\in L^2(V;\mathbb C^3)  $ in $L^2$-norm, we then have the $L^2$-convergence (mean square convergence) of the Fourier inversion formulae:
\begin{align}\Delta_s(\bm r)\colonequals {}&-\frac1{(2\pi)^3}\Fint_{\mathbb R^3}\frac{(\bm u\cdot\bm q)(\bm v\cdot\bm q)\widetilde {\bm E}_s(\bm q)}{|\bm q|^2}e^{-i\bm q\cdot \bm r}\D^3\bm q\notag\\{}&+\frac1{(2\pi)^3}\Fint_{\mathbb R^3}\frac{(\bm u\cdot\bm q)(\bm v\cdot\bm q)\widetilde {\bm E}(\bm q)}{|\bm q|^2}e^{-i\bm q\cdot \bm r}\D^3\bm q\overset{L^2}{\longrightarrow}\mathbf0\end{align}
as is evident from the bound estimate of the residual error based on  Parseval--Plancherel identity  $ \Vert\bm F\Vert^2_{L^2(\mathbb R^3;\mathbb C^3)}\colonequals\langle \bm F,\bm F\rangle_{\mathbb R^3}=\frac{1}{(2\pi )^3}\langle \widetilde {\bm F},\widetilde {\bm F}\rangle_{\mathbb R^3}\equalscolon\frac{1}{(2\pi )^3}\Vert\widetilde {\bm F}\Vert^2_{L^2(\mathbb R^3;\mathbb C^3)}$  \cite[p.~104]{Grafakos}:\begin{align}&\Vert\Delta_s\Vert_{L^2(\mathbb R^3;\mathbb C^3)}^2=\frac1{(2\pi)^3}\iiint_{\mathbb R^3}\left|\frac{(\bm u\cdot\bm q)(\bm v\cdot\bm q)[\widetilde {\bm E}_s(\bm q)-\widetilde {\bm E}(\bm q)]}{|\bm q|^2} \right\vert^2\D^3\bm r\notag\\\leq{}&\frac1{(2\pi)^3}\iiint_{\mathbb R^3}|\widetilde {\bm E}_s(\bm q)-\widetilde {\bm E}(\bm q) \vert^2\D^3\bm r=\Vert\bm E_s-\bm E\Vert_{L^2(V;\mathbb C^3)}^2\to0.\end{align}
Meanwhile, back in the  $\bm r $-space, we also have $\Vert\hat{\mathscr D}_{\bm u,\bm v}\bm E_s-\hat{\mathscr D}_{\bm u,\bm v}\bm E\Vert_{L^2(\mathbb R^3;\mathbb C^3)}^2\leq\sum_{|\bm\mu|=2}\Vert D^{\bm \mu} \hat{\mathscr N}(\bm E_s-\bm E)\Vert^2_{L^{2}(\mathbb R^3;\mathbb C^3)}=\Vert \bm E_s-\bm E\Vert^2_{L^{2}(V;\mathbb C^3)}\to0$, where the last equality comes from  a familiar identity  $\sum_{|\bm\mu|=2}\Vert D^{\bm \mu} \hat{\mathscr N}\bm F\Vert^2_{L^{2}(\mathbb R^3;\mathbb C^3)}=\Vert \bm F\Vert^2_{L^{2}(V;\mathbb C^3)}$ (see \cite[Theorem 9.9]{Gilbarg}, also \cite[Theorem 10.1.1]{Jost}).
 Therefore, the Fourier inversion formula \eqref{eq:FTinvN} is preserved as we take $L^2$-limits, and thus extends to all $\bm E\in  L^2(V;\mathbb C^3)$, just as claimed.\item By \eqref{eq:Npot_deriv_FT}, we have \begin{align}
((\hat{\mathscr G}-\hat\gamma)\bm E)(\bm r)+\bm E(\bm r)=-\frac1{(2\pi)^3}\Fint_{\mathbb R^3}\frac{\bm q\times[\bm q\times\widetilde{\bm E}(\bm q)]}{|\bm q|^2}e^{-i\bm q\cdot \bm r}\D^3\bm q,
\end{align}with $\widetilde{\bm E}(\bm q)=\iiint_V\bm E(\bm r)e^{i\bm q\cdot \bm r}\D^3\bm r$ for all $ \bm q\in\mathbb R^3$. Using the Parseval--Plancherel identity   $ \langle \bm F,\bm G\rangle_{\mathbb R^3}=\frac{1}{(2\pi )^3}\langle \widetilde {\bm F},\widetilde {\bm G}\rangle_{\mathbb R^3}$  \cite[p.~104]{Grafakos}, we may  compute that
\begin{align}\langle (\hat \gamma-\hat {\mathscr G})\bm E,\bm E\rangle_{V}=\langle\bm E,  (\hat \gamma-\hat {\mathscr G})\bm E\rangle_{V}=\frac1{(2\pi)^3}\iiint_{\mathbb R^3}\frac{|\bm q\cdot\widetilde{\bm E}(\bm q)|^{2}}{|\bm q|^2}\D^3\bm q,\end{align}which immediately leads to the inequality\begin{align}0\leq\langle  (\hat \gamma-\hat {\mathscr G})\bm E,\bm E\rangle_{V}\leq\frac1{(2\pi)^3}\iiint_{\mathbb R^3}|\widetilde{\bm E}(\bm q)|^{2}\D^3\bm q=\langle\bm E, \bm E\rangle_V,\end{align}an equivalent form of   \eqref{eq:Hermitian_ineq}.\end{enumerate} \end{proof}
\begin{proposition}[Fourier analysis of $ \hat{\mathscr G}$]\label{prop:Green_FT} \begin{enumerate}[leftmargin=*,  label=\emph{(\alph*)},ref=(\alph*),
widest=a, align=left]
\item We have the Fourier inversion formula:\begin{align}\label{eq:epsBorn}&(\bm u\cdot\nabla)(\bm v\cdot\nabla)\iiint_V\frac{\bm E(\bm r' )e^{-i(k-i\varepsilon)|\bm r-\bm r' |}}{4\pi|\bm r-\bm r' |}  \D^3 \bm r'\notag\\={}&-\frac1{(2\pi)^3}\Fint_{\mathbb R^3}\frac{(\bm u\cdot\bm q)(\bm v\cdot\bm q)\widetilde {\bm E}(\bm q)}{|\bm q|^2-(k-i\varepsilon)^2}e^{-i\bm q\cdot \bm r}\D^3\bm q,\quad\forall \bm u,\bm v\in\{\bm e_x,\bm e_y,\bm e_z\},\end{align}for  every $\varepsilon>0$ and $\bm E\in L^2(V;\mathbb C^3)$.
\item
The Born operator  $\hat{\mathscr B}=\hat I-\chi\hat {\mathscr G}\colon L^2(V;\mathbb C^3)\longrightarrow L^2(V;\mathbb C^3) $ has a limit representation:
 \begin{align}(\hat{\mathscr B}\bm E)(\bm r)=(1+\chi)\bm E(\bm r)+\chi\lim_{\varepsilon\to0^+}\frac1{(2\pi)^3}\Fint_{\mathbb R^3}\frac{\bm q\times[\bm q\times\widetilde{\bm E}(\bm q)]}{|\bm q|^2-(k-i\varepsilon)^2}e^{-i\bm q\cdot \bm r}\D^3\bm q. \label{eq:L2Born}\end{align}\end{enumerate}\end{proposition}
\begin{proof}
\begin{enumerate}[leftmargin=*,  label={(\alph*)},ref=(\alph*),
widest=a, align=left]
\item To prove \eqref{eq:epsBorn}, we may first consider $ \bm F\in C^\infty_0(V;\mathbb C^3)$. In this case, owing to the exponential decay factor $e^{-\varepsilon|\bm r-\bm r'|} $ in the convolution kernel and the vanishing modulus $|\bm F(\bm r)| $ at the boundary $\partial V$, the convolution $\iiint_V(4\pi|\bm r-\bm r' |)^{-1} \bm F(\bm r' )e^{-i(k-i\varepsilon)|\bm r-\bm r' |}\D^3 \bm r'$ yields a vector field of type $L^1(\mathbb R^3;\mathbb C^3)\cap C^2(\mathbb R^3;\mathbb C^3)$.  Thus, we may evaluate the Fourier transform of the convolution via an absolutely convergent integral and apply the Fubini theorem to interchange the order of integrations:\begin{align}&\iiint_{\mathbb R^3}\left[\iiint_V\frac{\bm F(\bm r' )e^{-i(k-i\varepsilon)|\bm r-\bm r' |}}{4\pi|\bm r-\bm r' |}  \D^3 \bm r'\right]e^{i\bm q\cdot \bm r}\D^3\bm r\notag\\\colonequals {}&\iiint_V \bm F(\bm r' )e^{i\bm q\cdot \bm r'}\D^3\bm r'\iiint_{\mathbb R^3}\frac{e^{-i(k-i\varepsilon)|\bm r'' |}}{4\pi|\bm r'' |}e^{i\bm q\cdot \bm r''}\D^3\bm r''=\frac{\widetilde {\bm F}(\bm q)}{|\bm q|^2-(k-i\varepsilon)^2},\end{align}and obtain the Fourier inversion formula in the form of a usual integral for every $  \bm F\in C^\infty_0(V;\mathbb C^3)$:\begin{align}\iiint_V\frac{\bm F(\bm r' )e^{-i(k-i\varepsilon)|\bm r-\bm r' |}}{4\pi|\bm r-\bm r' |}  \D^3 \bm r'=-\frac1{(2\pi)^3}\iiint_{\mathbb R^3}\frac{\widetilde {\bm F}(\bm q)}{|\bm q|^2-(k-i\varepsilon)^2}e^{-i\bm q\cdot \bm r}\D^3\bm q.\end{align}Now that we have the absolute convergence of the following integrals for all  $  \bm u,\bm v\in\{\bm e_x,\bm e_y,\bm e_z\}$: \begin{align}\iiint_{\mathbb R^3}\left\vert\frac{(\bm u\cdot\bm q)\widetilde{ \bm F} (\bm q)}{|\bm q|^2-(k-i\varepsilon)^2}\right\vert\D^3\bm q<+\infty,\;\iiint_{\mathbb R^3}\left\vert\frac{(\bm u\cdot\bm q)(\bm v\cdot\bm q)\widetilde{ \bm F} (\bm q)}{|\bm q|^2-(k-i\varepsilon)^2}\right\vert\D^3\bm q<+\infty, \end{align} which implies the continuity of corresponding Fourier integrals, we can proceed to confirm that for every  $\bm F\in C^\infty_0(V;\mathbb C^3)$, the following identity holds for all  $  \bm u,\bm v\in\{\bm e_x,\bm e_y,\bm e_z\}$:
\begin{align}(\hat{\mathscr D}^{(\varepsilon)}_{\bm u,\bm v}\bm F)(\bm r)\colonequals{}& (\bm u\cdot\nabla)(\bm v\cdot\nabla)\iiint_V\frac{\bm F(\bm r' )e^{-i(k-i\varepsilon)|\bm r-\bm r' |}}{4\pi|\bm r-\bm r' |}  \D^3 \bm r'\notag\\={}&-\frac1{(2\pi)^3}\iiint_{\mathbb R^3}\frac{(\bm u\cdot\bm q)(\bm v\cdot\bm q)\widetilde {\bm F}(\bm q)}{|\bm q|^2-(k-i\varepsilon)^2}e^{-i\bm q\cdot \bm r}\D^3\bm q.\end{align}
To deduce  \eqref{eq:epsBorn} from the equation above, it would suffice to prove that  both sides of  \eqref{eq:epsBorn} indeed define a continuous mapping from  $L^2(V;\mathbb C^3)$ to $ L^2(\mathbb R^3;\mathbb C^3) $.

Here, it is clear that the right-hand side of \eqref{eq:epsBorn} represents a continuous mapping from $L^2(V;\mathbb C^3) $ to $L^2(\mathbb R^3;\mathbb C^3) $, as the following estimate holds for every $\bm E\in L^2(V;\mathbb C^3) $:\begin{align}\frac
1{(2\pi)^3}\iiint_{\mathbb R^3}\left\vert\frac{(\bm u\cdot\bm q)(\bm v\cdot\bm q)\widetilde {\bm E}(\bm q)}{|\bm q|^2-(k-i\varepsilon)^2}\right\vert^2\D^3\bm q\leq\Vert\bm E\Vert_{L^2(V;\mathbb C^3)}\sup_{\bm q\in \mathbb R^3}\left\vert\frac{(\bm u\cdot\bm q)(\bm v\cdot\bm q)}{|\bm q|^2-(k-i\varepsilon)^2}\right\vert^2,\end{align}where the right-hand side is a finite multiple of  $\Vert\bm E\Vert_{L^2(V;\mathbb C^3)}$.

The left-hand side of \eqref{eq:epsBorn} represents a continuous mapping from $L^2(V;\mathbb C^3) $ to $L^2(\mathbb R^3;\mathbb C^3) $ as well, for two reasons.

First, for every $\bm E\in L^2(V;\mathbb C^3)$, one naturally has
\begin{align}&(\hat{\mathscr D}^{(\varepsilon)}_{\bm u,\bm v}\bm E)(\bm r)\notag\\={}&(\bm u\cdot\nabla)(\bm v\cdot\nabla)\iiint_{\mathbb R^3}\frac{\bm E(\bm r' )}{4\pi|\bm r-\bm r' |}  \D^3 \bm r'\notag\\{}&+(\bm u\cdot\nabla)(\bm v\cdot\nabla)\iiint_V\frac{\bm E(\bm r' )[e^{-i(k-i\varepsilon)|\bm r-\bm r' |}-1]}{4\pi|\bm r-\bm r' |}  \D^3 \bm r'.\end{align}Pick $R_V\colonequals \min_{\bm r\in\mathbb R^3}\max_{\bm r'\in V\cup\partial V}|\bm r'-\bm r| $ to be the radius of the circumscribing sphere, and we may assume, without loss of generality, that $V\subset O(\mathbf0,2R_V) $, where $ O(\bm r_0,\varepsilon)$ stands for an open ball centered at $\bm r_0$, with radius $\varepsilon$. Thus, we  have  the following bound estimate on the open ball $O(\mathbf0,2R_V) $
 \begin{align}&\Vert\hat{\mathscr D}^{(\varepsilon)}_{\bm u,\bm v}\bm E\Vert_{L^2(O(\mathbf0,2R_V),\mathbb C^3)}\notag\\\leq{}&\left(1+\sqrt{\iiint_{O(\mathbf0,2R_V)}\left\{ \iiint_V \Tr\left[\mathbf H_\varepsilon^{*}(\bm r,\bm r')\mathbf H_\varepsilon(\bm r,\bm r')\right]\D^3\bm r'\right\}\D^3\bm r}\right)\Vert\bm E\Vert_{L^2(V,\mathbb C^3)},\end{align}for the Hessian matrix $
\mathbf H_\varepsilon(\bm r,\bm r')\colonequals\nabla\nabla\frac{e^{-i(k-i\varepsilon)|\bm r-\bm r' |}-1}{4\pi|\bm r-\bm r' |}$ whose conjugate transpose  is $\mathbf  H_\varepsilon^{*}(\bm r,\bm r')$. Here, the multiple integral under the square root is finite, owing to the square integrability of the weak singularity of order $ O(|\bm r-\bm r'|^{-1})$.

Second, we note that so long as $\bm r\in \mathbb R^3\smallsetminus(V\cup\partial V) $, we have the square-integrable convolution kernel:\begin{align}\iiint_V\left\vert(\bm u\cdot\nabla)(\bm v\cdot\nabla)\frac{e^{-i(k-i\varepsilon)|\bm r-\bm r' |}}{4\pi|\bm r-\bm r' |}\right\vert^ 2 \D^3 \bm r<+\infty,\end{align}so we may differentiate under the integral sign for
$\hat{\mathscr D}^{(\varepsilon)}_{\bm u,\bm v}\bm E\in C^0(\mathbb R^3\smallsetminus(V\cup\partial V);\mathbb C^3)$ to deduce \begin{align}(\hat{\mathscr D}^{(\varepsilon)}_{\bm u,\bm v}\bm E)(\bm r)\colonequals \iiint_V\bm E(\bm r' )(\bm u\cdot\nabla)(\bm v\cdot\nabla)\frac{e^{-i(k-i\varepsilon)|\bm r-\bm r' |}}{4\pi|\bm r-\bm r' |}  \D^3 \bm r',\quad\bm r\in \mathbb R^3\smallsetminus(V\cup\partial V)\end{align}and perform the estimate
\begin{align}&\Vert\hat{\mathscr D}^{(\varepsilon)}_{\bm u,\bm v}\bm E\Vert_{L^2( \mathbb R^3\smallsetminus O(\mathbf0,2R_V),\mathbb C^3)}\notag\\\leq{}&\sqrt{\iiint_{ \mathbb R^3\smallsetminus O(\mathbf0,2R_V)}\left\{ \iiint_V \Tr\left[\mathbf H_\varepsilon^{*}(\bm r,\bm r')\mathbf H_\varepsilon(\bm r,\bm r')\right]\D^3\bm r'\right\}\D^3\bm r}\Vert\bm E\Vert_{L^2(V,\mathbb C^3)}.\end{align}
Here, the multiple integral under the square root is finite, owing  to the exponential decay factor $e^{-\varepsilon|\bm r-\bm r'|} $.

Combining the efforts above, we have verified the Fourier inversion formula as claimed in  \eqref{eq:epsBorn}.\item For any $ \varepsilon>0$, we can define the approximate Born operator $\hat{\mathscr B}^{(\varepsilon)}\colon L^2(V;\mathbb C^3)\longrightarrow L^2(V;\mathbb C^3)$ as \begin{align}(\hat{\mathscr B}^{(\varepsilon)}\bm E)(\bm r)\colonequals {}&(1+\chi)\bm E(\bm r)+\chi\frac1{(2\pi)^3}\Fint_{\mathbb R^3}\frac{\bm q\times[\bm q\times\widetilde{\bm E}(\bm q)]}{|\bm q|^2-(k-i\varepsilon)^2}e^{-i\bm q\cdot \bm r}\D^3\bm q\notag\\={}&(1+\chi)\bm E(\bm r)-\chi\nabla\times \nabla\times \iiint_V\frac{\bm E(\bm r' )e^{-i(k-i\varepsilon)|\bm r-\bm r' |}}{4\pi|\bm r-\bm r' |}  \D^3 \bm r'\notag\\={}&(1+\chi)\bm E(\bm r)+\chi\frac1{(2\pi)^3}\Fint_{\mathbb R^3}\frac{\bm q\times[\bm q\times\widetilde{\bm E}(\bm q)]}{|\bm q|^2}e^{-i\bm q\cdot \bm r}\D^3\bm q\notag\\&-\chi\nabla\times \nabla\times\iiint_V\frac{[e^{-i(k-i\varepsilon)|\bm r-\bm r' |}-1+i(k-i\varepsilon)|\bm r-\bm r' |]}{4\pi|\bm r-\bm r' |}  \bm E(\bm r')\D^3 \bm r'\end{align}where the equalities hold because of   \eqref{eq:epsBorn}. It is thus clear that $\hat{\mathscr B}^{(\varepsilon)}\colon L^2(V;\mathbb C^3)\longrightarrow L^2(V;\mathbb C^3)$ is a bounded linear operator for every  $ \varepsilon>0$, and the operator norm of \begin{align}&(\hat{\mathscr B}^{(\varepsilon)}\bm E)(\bm r)-(\hat{\mathscr B}\bm E)(\bm r)\notag\\={}&-\chi\nabla\times \nabla\times\iiint_V\frac{\bm E(\bm r' )[e^{-i(k-i\varepsilon)|\bm r-\bm r' |}-e^{-ik|\bm r-\bm r' |}+\varepsilon|\bm r-\bm r' |]}{4\pi|\bm r-\bm r' |}  \D^3 \bm r'\end{align}satisfies $\Vert\hat{\mathscr B}^{(\varepsilon)} -\hat{\mathscr B}\Vert_{L^2(V;\mathbb C^3)}\leq C_{k;V}\varepsilon^2 $ for some constant $C_{k;V} $ only dependent on the wave number $k$ and the shape of the scattering medium $V$, which confirms the  representation as claimed.\end{enumerate} \end{proof}\begin{remark}In \cite{Holt}, A.~R.~Holt \textit{et al.}~have recast the Born equation  with the dyadic Green function $(1+k^{-2}\nabla\nabla)[e^{-ik|\bm r-\bm r'|}/(4\pi|\bm r-\bm r'|)] $, and have obtained the Fourier transform of the Born equation in dyadic form \cite[(16)]{Holt}, which is mathematically equivalent to \eqref{eq:L2Born}.\eor
\end{remark}
 \subsection{\label{subsec:uniq_L2}Uniqueness theorem and scattering cross-section for $L^2$ fields}

\begin{theorem}[Generalized optical theorem for scattering cross-sections]\label{thm:GOT}The Born operator $\hat{\mathscr B}=\hat I-\chi\hat {\mathscr G}\colon L^2(V;\mathbb C^3)\longrightarrow L^2(V;\mathbb C^3) $ satisfies the following functional relation: \begin{align}&
\frac{|\chi|^2 k^5}{16\pi^2 } \varoiint_{|\bm n|=1}\left[\bm n\times\iiint_V \bm E(\bm r' ) e^{ik\bm n \cdot \bm r' }  \D^3\bm r' \right]\cdot\left[\bm n\times\iiint_V \bm E^*(\bm r' ) e^{-ik\bm n \cdot \bm r' }  \D^3\bm r' \right]\D\Omega\notag\\=&\I\left[\chi k^2\iiint_V|\bm E(\bm r )|^{2}\D^3\bm r \right]-\I\left[\chi k^2\iiint_V(\hat{\mathscr  B}\bm E)^*(\bm r)\cdot\bm E(\bm r )\D^3\bm r\right],\label{eq:GOT1}
\end{align}where $ \D\Omega=\sin\theta\D\theta\D\phi$ stands for the surface element on the unit sphere. \end{theorem}\begin{proof}We directly  evaluate the scalar product via \eqref{eq:L2Born} in Proposition~\ref{prop:Green_FT}:\begin{align}{}&\I\langle\bm E-\hat{\mathscr  B}\bm E,\chi k^2\bm E\rangle_{V}=\lim_{\varepsilon\to 0^+}\I\langle\bm E-\hat{\mathscr  B}^{(\varepsilon)}\bm E,\chi k^2\bm E\rangle_{V}\notag\\={}&-\I\left\{\chi k^2\iiint_V\left[\chi\lim_{\varepsilon\to0^+}\frac1{(2\pi)^3}\Fint_{\mathbb R^3}\frac{\bm q\times[\bm q\times\widetilde{\bm E}(\bm q)]}{|\bm q|^2-(k-i\varepsilon)^2}e^{-i\bm q\cdot \bm r}\D^3\bm q \right]^*\cdot\bm E(\bm r)\D^3\bm r\right\}.\end{align}Now, using the Parseval--Plancherel identity    $ \langle \bm F,\bm G\rangle_{\mathbb R^3}=\frac{1}{(2\pi )^3}\langle \widetilde {\bm F},\widetilde {\bm G}\rangle_{\mathbb R^3}$  \cite[p.~104]{Grafakos}, we obtain
\begin{align}\I\langle\bm E-\hat{\mathscr  B}\bm E,\chi k^2\bm E\rangle_{V}=-\frac{|\chi|^2k^2}{(2\pi)^3}\I\lim_{\varepsilon\to0^+}\iiint_{\mathbb R^3}\frac{\bm q\times[\bm q\times\widetilde{\bm E}^*(\bm q)]}{(|\bm q|-k-i\varepsilon)(|\bm q|+k+i\varepsilon) }\cdot\widetilde{\bm E}(\bm q)\D^3\bm q.\end{align}With the vector identity $\widetilde{\bm E}(\bm q)\cdot\{\bm q\times[\bm q\times\widetilde{\bm E}^*(\bm q)]\}=-[\bm q\times\widetilde{\bm E}(\bm q)]\cdot[\bm q\times\widetilde{\bm E}^*(\bm q)]=-|\bm q\times\widetilde{\bm E}(\bm q)|^{2} $, and the Plemelj jump relation
\begin{align}\I\lim_{\varepsilon\to0^+}\int_0^{+\infty}\frac{f(q)\D q }{q-k-i\varepsilon }=\pi f(k),\end{align}we arrive at
the generalized optical theorem\begin{align}
\I\langle\bm E-\hat{\mathscr  B}\bm E,\chi k^2\bm E\rangle_{V}={}&\frac{|\chi|^2k^2}{(2\pi)^3}\I\lim_{\varepsilon\to0^+}\int_0^{+\infty}|\bm q|^2\D|\bm q|\varoiint\D\Omega\frac{|\bm q\times\widetilde{\bm E}(\bm q)|^{2}}{|\bm q|^2-(k-i\varepsilon)^2 }\notag\\={}&\frac{|\chi|^2 k^5}{16\pi^2 }\varoiint_{|\bm n|=1}\left\vert\bm n\times\widetilde{\bm E}(k\bm n)\right\vert^2\D\Omega,\label{eq:GOT2}\end{align}for every $ \bm E\in L^2(V;\mathbb C^3)$. \end{proof}

Now, we are ready to work out the ``uniqueness theorem'' for the Born equation posed on the Hilbert space $ L^2(V;\mathbb C^3)$. For physically meaningful electromagnetic scattering problems related to a homogeneous non-accretive medium, we require that the susceptibility $\chi $ satisfy $ \I\chi\leq0$ in the Born equation \eqref{eq:L2Born}. When $\chi=0$, the Born operator trivializes to the identity operator.
In the following corollary, we will establish the uniqueness theorem for the physically meaningful and non-trivial susceptibilities $ \I\chi\leq0,\chi\neq0$, barring a special case $\chi=-1$.\begin{corollary}[Uniqueness criteria for Born equation on $ L^2(V;\mathbb C^3)$]\label{thm:uniq}For the Born operator\linebreak $\hat{\mathscr B}\colon L^2(V;\mathbb C^3)\longrightarrow L^2(V;\mathbb C^3) $ [as defined in \eqref{eq:L2Born}], the homogeneous equation $\Vert \hat{\mathscr B}\bm E\Vert_{{L^2(V;\mathbb C^3)}}=0$ admits only a trivial solution $\Vert\bm E\Vert_{L^2(V;\mathbb C^3)}=0 $ if  one of the following two conditions holds:\begin{enumerate}[leftmargin=*,  label=\emph{(\Alph*)},
widest=a, align=left]
\item
The dielectric medium is dissipative with $\I\chi<0$;
\item
The dielectric medium is non-dissipative with $\I\chi=0,\chi\neq0,-1$,  its boundary $ \partial V$ occupies zero volume $\iiint_{\partial V}\D^3\bm r=0 $, and the exterior volume $\mathbb R^3\smallsetminus(V\cup\partial V)$ is connected.

\end{enumerate}
\end{corollary}\begin{proof}For situation (A), we  substitute the condition $\Vert \hat{\mathscr B}\bm E\Vert_{{L^2(V;\mathbb C^3)}}=0  $ into the generalized optical theorem (Theorem \ref{thm:GOT}), and obtain \begin{align}&
\frac{|\chi|^2 k^5}{16\pi^2 } \varoiint_{|\bm n|=1}\left|\bm n\times\iiint_V \bm E(\bm r' ) e^{ik\bm n \cdot \bm r' }  \D^3\bm r' \right|^2\D\Omega=\I\left[\chi k^2\iiint_V|\bm E(\bm r )|^{2}\D^3\bm r \right].\label{eq:uniqthm}\end{align}
Now that the left-hand side of \eqref{eq:uniqthm} is non-negative, and the right-hand side of \eqref{eq:uniqthm} is non-positive, the only possibility is that both sides vanish. As $\I\chi\neq0 $, the vanishing right-hand side of \eqref{eq:uniqthm} implies $\Vert\bm E\Vert_{L^2(V;\mathbb C^3)}=0$.

For situation (B), we point out that  $\Vert \hat{\mathscr B}\bm E\Vert_{{L^2(V;\mathbb C^3)}}=0$  and  Theorem~\ref{thm:GOT} together imply $\bm q\times[\bm q\times\widetilde{\bm E}(\bm q)]=\mathbf 0 $ for all $|\bm q|=k$. As we further note that  $ \widetilde {\bm E}\in C^1(\mathbb R^3;\mathbb C^3)$ arises from the boundedness of second-order moments of $\bm E\in L^2(V;\mathbb C^3)$, we see that  $\bm q\times[\bm q\times\widetilde{\bm E}(\bm q)]/(|\bm q|^2-k^2) $ is bounded and continuous for all $\bm q\in \mathbb R^3$. Consequently, we can use the dominated convergence theorem to prove that \begin{align}\lim_{\varepsilon\to0^+}\iiint_{\mathbb R^3}\left|\frac{\bm q\times[\bm q\times\widetilde{\bm E}(\bm q)]}{|\bm q|^2-(k-i\varepsilon)^2}-\frac{\bm q\times[\bm q\times\widetilde{\bm E}(\bm q)]}{|\bm q|^2-k^2} \right\vert^2\D^3\bm q=0,\end{align}
so
\begin{align}\chi\frac1{(2\pi)^3}\Fint_{\mathbb R^3}\frac{\bm q\times[\bm q\times\widetilde{\bm E}(\bm q)]}{|\bm q|^2-k^2}e^{-i\bm q\cdot \bm r}\D^3\bm q\end{align}is a valid Fourier inversion formula for the corresponding term in the Born equation.
 Inside the dielectric volume $V$, we have\begin{align}(\hat{\mathscr B}\bm E)(\bm r)\colonequals (1+\chi)\bm E(\bm r)+\chi\frac1{(2\pi)^3}\Fint_{\mathbb R^3}\frac{\bm q\times[\bm q\times\widetilde{\bm E}(\bm q)]}{|\bm q|^2-k^2}e^{-i\bm q\cdot \bm r}\D^3\bm q=\mathbf0,\quad\forall \bm r\in V\smallsetminus V_0, \end{align}for some subset $V_0\subset V $ with zero volume $\iiint_{V_0}\D^3\bm r=0 $.
Outside the dielectric volume $V $, we  have
\begin{align}&\nabla\times \nabla\times\iiint_V\frac{\bm E(\bm r' )e^{-ik|\bm r-\bm r' |}}{4\pi|\bm r-\bm r' |}  \D^3 \bm r'\notag\\={}&-\frac1{(2\pi)^3}\Fint_{\mathbb R^3}\frac{\bm q\times[\bm q\times\widetilde{\bm E}(\bm q)]}{|\bm q|^2-k^2}e^{-i\bm q\cdot \bm r}\D^3\bm q=\mathbf 0,\quad\forall \bm r\in\mathbb R^3\smallsetminus(V\cup\partial V\cup V_0')\end{align}
for some subset $V_0'\subset  \mathbb R^3\smallsetminus(V\cup\partial V) $ with zero volume $\iiint_{V_0'}\D^3\bm r=0$, provided that $ \mathbb R^3\smallsetminus(V\cup\partial V)$ is connected.
This is because the vanishing scattering amplitudes in all directions\begin{align}\label{eq:van_sc_amp}\bm f(\bm n)\colonequals -\frac{\chi k^2}{4\pi} \bm n\times\left[\bm n\times\iiint_V\bm E(\bm r' )e^{ik\bm n\cdot \bm r'}\D^3\bm r'\right]\equiv\mathbf{0},\quad |\bm n|=1\end{align}
 imply
a vanishing scattering field in the connected open set $\mathbb R^3\smallsetminus(V\cup\partial V)$\begin{align}\bm E_\mathrm{sc}(\bm r)\colonequals \chi\nabla\times\nabla\times \iiint_V\frac{\bm E(\bm r' )e^{-ik|\bm r-\bm r' |}}{4\pi|\bm r-\bm r' |}  \D^3 \bm r'\equiv\mathbf{0},\quad\bm r\notin V\cup\partial V,\label{eq:Esc_def_int}\end{align}according to the unique continuation principle for the electric field $\bm E(\bm r)= \bm E_\mathrm{sc}(\bm r),\bm r\in\mathbb R^3\smallsetminus(V\cup\partial V)$   and the magnetic field  $\bm H(\bm r)\propto\nabla\times \bm E_\mathrm{sc}(\bm r),\bm r\in\mathbb R^3\smallsetminus(V\cup\partial V)$ that solve the Maxwell equations with the Silver--M\"uller radiation conditions   \cite[Corollary~4.10]{ColtonKress1983}. (Alternatively, one may invoke the unique continuation principle for the Helmholtz equation \cite[Theorem 8.6]{ColtonKress}.\footnote{Here, we have the transversality condition $ \nabla\cdot \bm E_\mathrm{sc}(\bm r)=0,\bm r\in\mathbb R^3\smallsetminus(V\cup\partial V)$ and the Helmholtz equation $ (\nabla^2+k^2)\bm E_\mathrm{sc}(\bm r)=\mathbf 0,\bm r\in\mathbb R^3\smallsetminus(V\cup\partial V)$ according to the defining  integral of $\bm E_\mathrm{sc}(\bm r),\bm r\in\mathbb R^3\smallsetminus(V\cup\partial V)$ in \eqref{eq:Esc_def_int}, so $  \bm r\cdot \bm E_\mathrm{sc}(\bm r),\bm r\in\mathbb R^3\smallsetminus(V\cup\partial V)$ is also annihilated  by the Helmholtz operator \cite[(9.110)]{Jackson:EM}, while both $ \bm E_\mathrm{sc}(\bm r)$ and $\nabla\times \bm E_\mathrm{sc}(\bm r) $  admit multipole expansions  \cite[(9.122)]{Jackson:EM} for sufficiently large $|\bm r|$. In particular, when $ V\subset O(\mathbf 0,R)$, we can choose $ |\bm r|>R$. Such multipole expansions vanish identically, as we study certain integrals involving \eqref{eq:van_sc_amp}  \cite[(9.122)]{Jackson:EM}. From $ \bm E_\mathrm{sc}(\bm r)=\mathbf 0,|\bm r|>R$, we can deduce $ \bm E_\mathrm{sc}(\bm r)=\mathbf0,\bm r\in\mathbb R^3\smallsetminus(V\cup\partial V)$ by unique continuation. })

Now, patching the identities inside and outside the volume $V$, we obtain
\begin{align}(1+\chi)\frac 1{(2\pi)^3}\Fint_{\mathbb R^3}\widetilde{\bm E}(\bm q)e^{-i\bm q\cdot \bm r}\D^3\bm q+\chi\frac1{(2\pi)^3}\Fint_{\mathbb R^3}\frac{\bm q\times[\bm q\times\widetilde{\bm E}(\bm q)]}{|\bm q|^2-k^2}e^{-i\bm q\cdot \bm r}\D^3\bm q=\mathbf0, \end{align} for all $ \bm r\in\mathbb R^3\smallsetminus(V_0\cup\partial V\cup V_0')$. Recalling  the  condition  $ \iiint_{\partial V}\D^3\bm r=0$,  we have $\iiint_{V_0\cup\partial V\cup V'_0}\D^3\bm r=0 $ as a consequence.
Therefore, by performing  Fourier transform on an identity that holds in $\bm r $-space with the possible exception of a subset $ V_0\cup\partial V\cup V_0'$ with zero volume, we may deduce \begin{align}(1+\chi)\widetilde{\bm E}(\bm q)+\chi\frac{\bm q\times[\bm q\times\widetilde{\bm E}(\bm q)]}{|\bm q|^2-k^2}=\mathbf0,\quad\forall \bm q\in\mathbb R^3\smallsetminus V_0''\label{eq:qspace}\end{align}where   $\iiint_{V_0''}\D^3\bm q=0$. Now, after multiplying  \eqref{eq:qspace} by $\bm q\times $, we obtain\begin{align}
(1+\chi)\bm q\times\widetilde{\bm E}(\bm q)-\chi|\bm q|^2\frac{\bm q\times\widetilde{\bm E}(\bm q)}{|\bm q|^2-k^2}=\mathbf0,\quad\forall \bm q\in\mathbb R^3\smallsetminus V_0'',\label{eq:qxE_0}\end{align}which implies $\bm q\times\widetilde{\bm E}(\bm q)=\mathbf 0,\forall \bm q\in\mathbb R^3\smallsetminus (V_0''\cup\partial O(\mathbf 0,|\sqrt{1+\chi}|k) )$. Referring back to  \eqref{eq:qspace}, and using the condition that $\chi\neq-1 $, we then arrive at \begin{align}\widetilde{\bm E}(\bm q)=\mathbf 0,\quad\forall \bm q\in\mathbb R^3\smallsetminus (V_0''\cup\partial O(\mathbf 0,|\sqrt{1+\chi}|k) )  \end{align}which implies $\Vert \bm E\Vert_{L^2(V;\mathbb C^3)}=(2\pi)^{-3/2}\Vert \widetilde {\bm E}\Vert_{L^2(\mathbb R^3;\mathbb C^3)}=0 $, just as claimed. \end{proof}\begin{remark}Generally speaking, the  condition   $\iiint_{\partial V}\D^3\bm r=0$ in situation (B) is not redundant, because there exist fractal structures  that are bounded open sets whose boundaries have positive ``volume'' (in terms of Lebesgue measure), while $\mathbb R^3\smallsetminus(V\cup\partial V) $ remains connected  \cite{Boerger}.\eor\end{remark}\begin{remark}The uniqueness criterion in the $\chi=-1$ case is much more intricate, yet still has important implications. This exceptional situation will be handled by Proposition~\ref{prop:sigma01}, as well as Propositions~\ref{prop:kerspace} and \ref{prop:smooth_bd}.\eor\end{remark}
\subsection{Existence theorem and spectral analysis of $L^2$ fields\label{subsec:exist_L2}}Unlike the finite-dimensional square matrices, the existence condition $ \ran\hat{\mathscr B}=L^2(V;\mathbb C^3)$ in the Born equation does not follow immediately from the uniqueness condition  $ \ker\hat{\mathscr B}=\{\mathbf0\}$. To bridge the gap between
the uniqueness theorem and the  existence theorem, we need to consider Fredholm operators \cite[Chap.~XI]{JConway}  on Hilbert spaces, which can be deemed as  ``well-behaved  matrices of infinite dimensions''.

A bounded linear operator
 $\hat{\mathscr O}\colon \mathfrak h\longrightarrow \mathfrak h $ acting on a Hilbert space $ \mathfrak h  $ is called a Fredholm operator if it has closed range
 $ \ran(\hat{\mathscr O})=\Cl(\ran(\hat{\mathscr O}))$, a finite-dimensional kernel space  $ \dim\ker(\hat{\mathscr O})<+\infty$, and a finite-dimensional co-kernel space   $ \dim[\ran(\hat{\mathscr O})]^\bot<+\infty$.

For a Fredholm operator $\hat{\mathscr O}:\mathfrak h\longrightarrow \mathfrak h $, its index is given by $\ind(\hat{\mathscr O})=\dim\ker(\hat{\mathscr O})-\dim[\ran(\hat{\mathscr O})]^\bot $. A zero-index Fredholm operator is has a bounded inverse if and only if its kernel space is trivial.

\begin{lemma}[Invertible operators associated with $ \hat {\mathscr G}-\hat \gamma$]For $\chi\notin(-\infty,-1] $, the inverse $[\hat I-\chi(\hat {\mathscr G}-\hat \gamma)]^{-1}\colon L^2(V;\mathbb C^3)\longrightarrow L^2(V;\mathbb C^3) $ is a well-defined bounded operator.\label{lm:posdef}\end{lemma}\begin{proof}In view of the inequality $ 0\leq\langle\bm E,(\hat\gamma-\hat{\mathscr G})\bm E \rangle_V\leq\langle\bm E,\bm E \rangle_V$ [see \eqref{eq:Hermitian_ineq}], we know that the Hermitian operator  $ \hat \gamma-\hat {\mathscr G}\colon L^2(V;\mathbb C^3)\longrightarrow L^2(\mathbb R^3;\mathbb C^3)$ is    positive-semidefinite, with operator norm $\Vert \hat \gamma-\hat {\mathscr G}\Vert_{L^2(\mathbb R^3;\mathbb C^3)}\leq1 $. We accordingly have an estimate of its spectrum as $ \sigma( \hat \gamma-\hat {\mathscr G})\subset [0,1]$.

Now, if $ \chi\notin(-\infty,-1]\cup\{0\}$, then $-\chi^{-1}\notin\sigma(\hat \gamma-\hat {\mathscr G}) $, and  the inverse of $\chi^{-1}\hat I+ (\hat \gamma-\hat {\mathscr G}) $ must be accordingly a bounded linear operator. The case $ \chi=0$ leads us to the identity operator.
  \end{proof}
\begin{proposition}[Existence criteria for Born equation on $ L^2(V;\mathbb C^3)$] \label{prop:Fredholm} If  $ \chi\notin(-\infty,-1]$, the Born operator $\hat{\mathscr B}=\hat I-\chi\hat {\mathscr G}\colon L^2(V;\mathbb C^3)\longrightarrow L^2(V;\mathbb C^3) $ is a Fredholm operator with index zero $\ind(\hat{\mathscr B})=0 $. In other words, the uniqueness theorem (a trivial kernel space $\ker(\hat{\mathscr B})=\ker(\hat I-\chi\hat {\mathscr G})=\{\mathbf0\} $) implies a bounded inverse $(\hat I-\chi\hat {\mathscr G} )^{-1}$ if  $ \chi\notin(-\infty,-1]$.
\end{proposition}
\begin{proof}
According to Lemma~\ref{lm:posdef}, when   $ \chi\notin(-\infty,-1]$, the operator $\hat I-\chi(\hat {\mathscr G}-\hat \gamma)\colon L^2(V;\mathbb C^3)\longrightarrow L^2(V;\mathbb C^3) $ is invertible, with $\ker[\hat I-\chi(\hat {\mathscr G}-\hat \gamma)] =\{\mathbf0\}$ and $\ran[\hat I-\chi(\hat {\mathscr G}-\hat \gamma)]=L^2(V;\mathbb C^3) $. Thus, we have $\ind[\hat I-\chi(\hat {\mathscr G}-\hat \gamma)]=0 $.

Meanwhile, the Hilbert--Schmidt operator $\hat {\mathscr \gamma}\colon L^2(V;\mathbb C^3)\longrightarrow L^2(V;\mathbb C^3)$ is necessarily  compact.

Thus, for    $ \chi\notin(-\infty,-1]$, we have established the Born operator as a zero-index Fredholm operator perturbed by a compact operator.
It is a fundamental result (see  \cite[p.~508]{METaylor}  or \cite[p.~366]{JConway}) that  compact perturbation turns a  Fredholm operator  into  a Fredholm operator with the index intact.  Hence, we have $\ind(\hat{\mathscr B}) =\ind[\hat I-\chi(\hat {\mathscr G}-\hat \gamma)-\chi\hat \gamma]=\ind[\hat I-\chi(\hat {\mathscr G}-\hat \gamma)]=0$, whenever $\chi\notin(-\infty,-1] $.
 \end{proof}

\begin{proposition}[Absence of  residual spectrum]So long as the volume $V$ is bounded and open, we have $\sigma_r(\hat{\mathscr G})=\varnothing $. \end{proposition}\begin{proof}Writing $ (\hat\gamma\bm E)(\bm r)=\iiint_{V}\hat \Gamma(\bm r,\bm r') \bm E(\bm r')\D^3\bm r$  for a convolution kernel\begin{align}
\hat \Gamma(\bm r,\bm r')\colonequals\frac{k^2e^{-ik|\bm r-\bm r' |} }{4\pi|\bm r-\bm r' |}\mathbf1+\nabla\nabla\frac{e^{-ik|\bm r-\bm r' |}-1}{4\pi|\bm r-\bm r' |}=\hat\Gamma(\bm r',\bm r),
\end{align}with $ \mathbf1$ being the $3\times3$ identity matrix, and $ \nabla\nabla f(\bm r)$ being the Hessian matrix of the function $ f(\bm r)$, we have   the following transpose  identity {\small\begin{align}&\langle \bm F,(\lambda\hat I-\hat{\mathscr  G})\bm E\rangle_{V}\notag\\={}&\langle \bm F,\lambda\bm E\rangle_{V}+\frac1{(2\pi)^3}\iiint_{\mathbb R^3}\frac{[\bm q\cdot\widetilde{\bm F}^*(\bm q)][\bm q\cdot\widetilde{\bm E}(\bm q)]}{|\bm q|^2}\D^3\bm q-\iiint_V\bm F^*(\bm r)\cdot\left[ \iiint_V\hat\Gamma(\bm r,\bm r') \bm E(\bm r')\D^3\bm r'\right]\D^3\bm r\notag\\={}&\langle \bm E^*,\lambda\bm F^*\rangle_{V}+\frac1{(2\pi)^3}\iiint_{\mathbb R^3}\frac{[\bm q\cdot\widetilde{\bm E}(\bm q)][\bm q\cdot\widetilde{\bm F}^*(\bm q)]}{|\bm q|^2}\D^3\bm q-\iiint_V\bm E(\bm r')\cdot\left[ \iiint_V\hat\Gamma(\bm r,\bm r') \bm F^{*}(\bm r)\D^3\bm r\right]\D^3\bm r'\notag\\={}&\langle\bm E^*,(\lambda\hat I-\hat{\mathscr  G})\bm F^*\rangle_{V}.\label{eq:Tid}\end{align}}In the derivation above, we have    interchanged the integrals with respect to $\D^3\bm r $ and $\D^3\bm r'$, justifiable by the absolute integrability of the iterated integrals (which derives from the Hilbert--Schmidt bound in $ \iiint_V\left\{\iiint_V\Tr[\hat {\Gamma}^*(\bm r,\bm r')\hat {\Gamma}(\bm r,\bm r')]\D^3 \bm r'\right\}\D^3\bm r<+\infty$) and an application of the  the Fubini theorem.

Now, suppose that the uniqueness theorem holds, \textit{i.e.}~$ \ker(\lambda\hat I-\hat{\mathscr G}) =\{\mathbf 0\}$. We point out that with the transpose identity \eqref{eq:Tid}, we may deduce that $\Cl(\ran(\lambda\hat I-\hat{\mathscr G}))=\Cl((\lambda\hat I-\hat{\mathscr G})L^{2}(V;\mathbb C^3))=L^{2}(V;\mathbb C^3)$, which shows that $\lambda\notin\sigma_r(\hat{\mathscr G}) $. If the opposite is true, we will be able to find a non-zero member of $L^{2}(V;\mathbb C^3) $ that is orthogonal to all the elements of $(\lambda\hat I-\hat{\mathscr G})L^{2}(V;\mathbb C^3) $, which  means that for some $\bm F\neq\mathbf{0} $, we have $ \langle \bm F,(\lambda\hat I-\hat{\mathscr G})\bm E\rangle=\langle\bm E^*,(\lambda\hat I-\hat{\mathscr G})\bm F^*\rangle=0,\forall \bm E\in L^{2}(V;\mathbb C^3) $. This immediately entails $ (\lambda\hat I-\hat{\mathscr G})\bm F^*=\mathbf{0}$. According to the  assumption that ~$ \ker(\lambda\hat I-\hat{\mathscr G}) =\{\mathbf 0\}$, we have $\bm F^*=\mathbf{0}$ as well. The contradiction means that we must have  $\Cl(\ran(\lambda\hat I-\hat{\mathscr G}))=L^{2}(V;\mathbb C^3)$.  \end{proof}
\begin{proposition}[Spectral r\^ole of $ \chi=-1$]\label{prop:sigma01}So long as the volume  $V$ is bounded and open, we have $-1\in\sigma_p(\hat{\mathscr G})$ as an eigenvalue of infinite multiplicity $\dim\ker(\hat I+\hat{\mathscr G})=+\infty $, and $0\in\sigma_p(\hat{\mathscr G})\cup\sigma_c(\hat{\mathscr G}) $ being outside the resolvent set $ \rho(\hat{\mathscr G})$.
\end{proposition}
\begin{proof}Take any $f\in C^\infty_0(V;\mathbb C)$, then it follows that
\begin{align}((\hat I+\hat{\mathscr G})\nabla f)(\bm r)={}&\nabla\times\nabla\times \iiint_V\frac{\nabla'f(\bm r' )e^{-ik|\bm r-\bm r' |}}{4\pi|\bm r-\bm r' |}  \D^3 \bm r'\notag\\={}&-\nabla\times\iiint_V\nabla'\times\left[\frac{\nabla'f(\bm r' )e^{-ik|\bm r-\bm r' |}}{4\pi|\bm r-\bm r' |}\right]  \D^3 \bm r'=\mathbf0,\end{align}where one can use the Gau{\ss} theorem to convert the last volume integral  into a surface integral with vanishing integrand. This proves that  $\dim\ker(\hat I+\hat{\mathscr G})=+\infty$ and $-1\in\sigma_p(\hat{\mathscr G})$.

If we assume that $0\in \rho(\hat{\mathscr G})$, then $\hat {\mathscr G}-\hat \gamma $ will become a Fredholm operator, satisfying $\dim\ker(\hat {\mathscr G}-\hat \gamma)<+\infty $, which contradicts the fact that
\begin{align}&((\hat {\mathscr G}-\hat \gamma)\nabla\times\bm F)(\bm r)=\nabla\left[ \nabla\cdot\iiint_V\frac{\nabla'\times\bm F(\bm r')}{4\pi|\bm r-\bm r'|} \D^3\bm r'\right]\notag\\={}&-\nabla\left[ \iiint_V\nabla'\cdot\left(\frac{\nabla'\times\bm F(\bm r')}{4\pi|\bm r-\bm r'|} \right)\D^3\bm r'\right]=\mathbf0,  \end{align}for all $  \bm F\in C^\infty_0(V;\mathbb C^3)$,
\textit{i.e.}~$\dim\ker(\hat {\mathscr G}-\hat \gamma)=+\infty$. As the residual spectrum is empty $\sigma_r(\hat{\mathscr G})=\varnothing $, we then must have $0\in\sigma_p(\hat{\mathscr G})\cup\sigma_c(\hat{\mathscr G})\cup\sigma_r(\hat{\mathscr G})=\sigma_p(\hat{\mathscr G})\cup\sigma_c(\hat{\mathscr G})$.
 \end{proof}
\begin{remark}The eigenvalue $-1\in\sigma_p(\hat{\mathscr G})$  of infinite multiplicity $\dim\ker(\hat I+\hat{\mathscr G})=+\infty $ seems to suggest that the light scattering problem becomes seriously ill-conditioned for $\chi=-1 $, reminiscent of the  postulated singular behavior at $ \epsilon_r=0$ in \cite{Budko,Budko2007}. However, we will show, in  Proposition~\ref{prop:smooth_bd}, that such a concern about the case of  $\chi=-1$ is physically irrelevant.\eor\end{remark}

\begin{proposition}[Spectral r\^oles of plasmonic permittivities]\label{prop:sigma_conn}If  $\mathbb R^3\smallsetminus(V\cup\partial V)$ is connected and $\iiint_{\partial V}\D^3\bm r=0 $, then  $ (-1,0]\subset \rho(\hat{\mathscr G})\cup\sigma_c(\hat{\mathscr G}) $. \end{proposition}
\begin{proof}For $-1<\lambda<0 $,  the uniqueness criterion in Corollary~\ref{thm:uniq}(B) implies that $\lambda$ is not an eigenvalue, so it must belong to the set $ \rho(\hat{\mathscr G})\cup\sigma_c(\hat{\mathscr G})\cup\sigma_r(\hat{\mathscr G})=\rho(\hat{\mathscr G})\cup\sigma_c(\hat{\mathscr G}) $.

For $\lambda=0$, we can still show that $\Vert \hat{\mathscr G}\bm E\Vert_{L^2(V;\mathbb C^3)}=0$ implies $ \Vert \bm E\Vert_{L^2(V;\mathbb C^3)}=0$, so $0\notin\sigma_p(\hat{\mathscr G}) $. To show this, we recall \eqref{eq:GOT2} in Theorem \ref{thm:GOT}, which
ensures that the relation   $\Vert \hat{\mathscr G}\bm E\Vert_{L^2(V;\mathbb C^3)}=0$ leads us to $\bm q\times\widetilde{\bm E}(\bm q)=\mathbf0,\forall |\bm q|=k$. Accordingly, we may adapt the proof of the uniqueness criterion in Corollary~\ref{thm:uniq}(B) to show that both
 \begin{align}\widetilde{\bm E}(\bm q)+\frac{\bm q\times[\bm q\times\widetilde{\bm E}(\bm q)]}{|\bm q|^2-k^2}=\mathbf0,\quad\forall \bm q\in\mathbb R^3\smallsetminus V_0''\end{align}and
 \begin{align} \bm q\times\widetilde{\bm E}(\bm q)-|\bm q|^2\frac{\bm q\times\widetilde{\bm E}(\bm q)}{|\bm q|^2-k^2}=\mathbf0,\quad\forall \bm q\in\mathbb R^3\smallsetminus V_0''\end{align}hold, where $\iiint_{V_0''}\D^3\bm q=0 $. Therefore, we may proceed to conclude
that $\Vert \bm E\Vert_{L^2(V;\mathbb C^3)}=\linebreak\frac{1}{(2\pi)^{3/2}} \Vert \widetilde{\bm E}\Vert_{L^2(\mathbb R^3;\mathbb C^3)}=0$.
 \end{proof}
\begin{remark}In classical electrodynamics, the plasmon range refers to the susceptibilities satisfying $ \chi<-1$. The proposition above suggests that the Born equation on $ L^2(V;\mathbb C^3)$ is potentially ill-posed for any point in the plasmon range, which hearkens back to the hypothesis on a continuum of resonance modes in \cite{Rahola}. In the next section, we show that the Hilbert space  $ L^2(V;\mathbb C^3)$ does not do justice to the optical resonance problem for materials in the plasmon range. By narrowing down to the physically meaningful  Hilbert subspace $ \Phi(V;\mathbb C^3)=\Cl(C^\infty(V;\mathbb C^3) \cap\ker(\nabla\cdot)\cap L^2(V;\mathbb C^3)  )$, the Born equation turns out to be  well-posed for $ \chi\in(-\infty,-2)\cup (-2,-1]$, provided that the dielectric boundary $ \partial V$ is smooth and the exterior volume $ \mathbb R^3\smallsetminus(V\cup\partial V)$ is connected.\eor\end{remark}

\section{Transversality condition and optical resonance theorem\label{sec:div0}}
\setcounter{equation}{0}
Motivated by one of the Maxwell equations $ \nabla\cdot\bm E=0$,  we construct a function space  $T(V;\mathbb C^3)= L^2(V;\mathbb C^3)\cap  C^\infty(V;\mathbb C^3) \cap\ker(\nabla\cdot)$ of square-integrable, smooth and divergence-free vector fields. We use the $ L^2$-closure of $ T(V;\mathbb C^3)$ to define
the Hilbert subspace $\Phi(V;\mathbb C^3)\colonequals \Cl(T(V;\mathbb C^3)) $.

Not only does the transversality condition enhance the uniqueness theorem (\S\ref{subsec:uniq_eps0}), it also leads to drastic improvement of the existence theorem (\S\ref{subsec:cpt_poly}) by ensuring polynomial compactness. In addition to showing that the quadratic monic polynomial $ \hat{\mathscr G}(\hat I+2\hat{\mathscr G})/2\colon \Phi(V;\mathbb C^3)\longrightarrow \Phi(V;\mathbb C^3)$ a minimal compact polynomial (\S\ref{subsec:cpt_poly}), we shall reveal the cubic polynomial  $ \hat{\mathscr G}(\hat I+2\hat{\mathscr G})^{2}\colon \Phi(V;\mathbb C^3)\longrightarrow \Phi(V;\mathbb C^3)$  as a Hilbert--Schmidt operator (\S\ref{subsec:HS_poly}).

In this section, we will  build physically meaningful uniqueness and existence theorems for electromagnetic scattering. These two results regarding the structure of the  \textit{physical spectrum }$\sigma^\Phi(\hat{\mathscr G})$  for \textit{smooth dielectrics} will depend on the following topological and geometric requirements:
\begin{enumerate}[leftmargin=*,  label=(\arabic*),
widest=a, align=left]
\item\label{itm:bd_open_V}
The dielectric volume $V=\bigcup_{j=1}^{N_V}V_j\Subset \mathbb R^3$ is a bounded open set with finitely many connected components $V_{1},\dots,V_{N_V}$  ($N_V<+\infty$);
 \item Each connected component $V_j$ has a smooth  boundary surface $\partial V_j $; \item\label{itm:disjoint_bdy} The dielectric boundary  $\partial V=\bigsqcup_{j=1}^{N_V}\partial V_j $ is the union of mutually disjoint subsets $\partial V_1$, $\dots$, $\partial V_{N_V}$ satisfying  $\partial V_j\cap\partial V_{j'}=\varnothing,1\leq j<j'\leq N_V$.
\end{enumerate}

Here, we say a few words about geometric smoothness. We  call a  subset $ \Sigma\subset \mathbb
R^3$ a  smooth surface if it is a  2-manifold of $ C^\infty$-class   \cite[p.~52]{doCarmo}.
For a bounded open set $V\Subset \mathbb R^3 $ with smooth boundary $\partial V $,   the \textit{outward}  unit normal $\bm n(\bm r) $ on the boundary surface $ \partial V$ [\textit{i.e.}~pointing from the ``inside'' $V\smallsetminus\partial V$ to the ``outside'' $\mathbb R^3\smallsetminus(V\cup \partial V)$] is well-defined --- ``single-sided surface'' anomalies like the M\"obius strip or the Klein bottle will not occur --- see \cite{Samelson}, also  \cite[p.~411]{METaylor},  \cite[p.~114]{doCarmo}, and \cite[p.~440]{Alexandroff}  for the orientability (two-sidedness) of compact surfaces.

The major tools in this section are boundary traces and fractional order Sobolev spaces (in particular, $ H^{1/2}$ and $ H^{-1/2}$), which  enable us to generalize certain vector calculus identities. For example, if a bounded open volume  $V\Subset\mathbb R^3$ has smooth boundary $ \partial V$,  and  $ \bm F\in L^{2}(V;\mathbb C^3)\cap C^1(V;\mathbb C^3)$, $ \nabla\cdot\bm F\in L^2(V;\mathbb C)$, $ f\in C^1(\mathbb R^3;\mathbb C)$, then the following generalized Green identity \cite[p.~206]{DautrayLionsVol3} holds:
\begin{align}\iiint_V\bm F^*(\bm r)\cdot\nabla f(\bm r)\D^3\bm r+\iiint_V f(\bm r)\nabla\cdot\bm F^*(\bm r)\D^3\bm r ={_{H^{-1/2}(\partial V;\mathbb C)}\langle {\bm n\cdot \bm F}|f\rangle_{H^{1/2}(\partial V;\mathbb C)}}.\end{align}Here, the last item  represents the canonical pairing  between the boundary traces $\bm n\cdot\bm  F\in H^{-1/2}(\partial V;\mathbb C)  $ and $f\in H^{1/2}(\partial V;\mathbb C)  $, and  it necessarily reduces to a usual Lebesgue integral \begin{align}_{H^{-1/2}(\partial V;\mathbb C)}\langle {\bm n\cdot \bm F}|f\rangle_{H^{1/2}(\partial V;\mathbb C)}=\varoiint_{\partial V}f(\bm r)\bm n\cdot\bm F^*(\bm r)\D S \label{eq:bd_tr_can_pair}\end{align}  if $\bm F$ is  smooth up to the boundary. For any bounded open set $M\Subset\mathbb R^3$ with smooth boundary surface $\Sigma=\partial M\subset\mathbb R^3  $, the Hilbert space $H^{1/2}(\Sigma;\mathbb C) $ is defined as $ H^{1/2}(\Sigma;\mathbb C)\colonequals \{ f:\Sigma\longrightarrow\mathbb C | \Vert f\Vert_{H^{1/2}(\Sigma;\mathbb C)}<+\infty\}$. Here, the  $H^{1/2} $-norm (also written as the $W^{1/2,2} $-norm \cite{Adams,HsiaoKleinman}) is given by\begin{align}\Vert f\Vert_{H^{1/2}(\Sigma;\mathbb C)}\colonequals \sqrt{\varoiint_\Sigma|f(\bm r)|^2\D S+\varoiint_\Sigma\left[\varoiint_\Sigma\frac{|f(\bm r)-f(\bm r')|^2}{|\bm r-\bm r'|^3}\D S'\right]\D S}.\end{align}
The space $H^{-1/2}(\Sigma;\mathbb C) $ is the Hilbert dual  of $H^{1/2}(\Sigma;\mathbb C) $, \textit{i.e.}~the totality of continuous linear functionals acting on  $H^{1/2}(\Sigma;\mathbb C)$ via the canonical pairing.

\subsection{Transversality condition and  uniqueness theorem for $ \epsilon_r=0$\label{subsec:uniq_eps0}}In \S\ref{subsec:uniq_L2}, we started our spectral analysis of $ L^2$ fields by showing that the Green operator $ \hat{\mathscr G}$ maps $ L^2(V;\mathbb C^3)$ to a subset of $ L^2(V;\mathbb C^3)$. Likewise, we will base the spectral analysis of transverse fields on the premise that  $ \hat{\mathscr G}$ maps $ \Phi(V;\mathbb C^3)$ to a subset of $ \Phi(V;\mathbb C^3)$.  \begin{lemma}[Green operator as endomorphism on $ \Phi(V;\mathbb C^3)$]\label{lm:Phi_endo}For any bounded and open volume $V$, the operator  $\hat {\mathscr G}\colon \Phi(V;\mathbb C^3)\longrightarrow \Phi(V;\mathbb C^3)$ is an endomorphism, satisfying  $\hat {\mathscr G}\Phi(V;\mathbb C^3)\subset \Phi(V;\mathbb C^3) $. \end{lemma}\begin{proof}First, we show that   $\hat {\mathscr G}(T(V;\mathbb C^3))\subset T(V;\mathbb C^3) $.

For any  $ \bm E\in L^2(V;\mathbb C^3)\cap  C^\infty(V;\mathbb C^3) $, we define \begin{align}(\hat{\mathscr C}\bm E)(\bm r)\colonequals \iiint_V\frac{\bm E(\bm r') e^{-ik|\bm r-\bm r' |}}{4\pi|\bm r-\bm r' |}\D^3 \bm r', \end{align}by an
extension of the notation in  \eqref{eq:Born'}.
We may use integration by parts to verify that  $\hat{\mathscr C}\bm E\in  L^2(V;\mathbb C^3)\cap C^\infty(V;\mathbb C^3)   $, and $(\nabla^2+k^2)(\hat{\mathscr C}\bm E)(\bm r)=-\bm E(\bm r),\forall\bm r\in V $. Therefore, we have    $\hat {\mathscr G}\bm E\in L^2(V;\mathbb C^3)\cap  C^\infty(V;\mathbb C^3)$. To demonstrate the transversality condition $\hat{\mathscr G}\bm E\in\ker(\nabla\cdot)  $, it would suffice to compute $\nabla\cdot \nabla\times\nabla\times(\hat{\mathscr C}\bm E)=0$.

Now that we have confirmed   $\hat {\mathscr G}(T(V;\mathbb C^3))\subset T(V;\mathbb C^3) $, and the continuity of $\hat {\mathscr G}: L^2(V;\mathbb C^3)\longrightarrow  L^2(V;\mathbb C^3) $ entails that $\hat {\mathscr G}(\Cl(\mathfrak A))\subset \Cl(\hat {\mathscr G}(\mathfrak A)) $ for any subset $\mathfrak A\subset L^2(V;\mathbb C^3)$, we have   $\hat {\mathscr G}\Cl(T(V;\mathbb C^3))\subset \Cl(\hat {\mathscr G}(T(V;\mathbb C^3)))\subset \Cl(T(V;\mathbb C^3))=\Phi(V;\mathbb C^3)$, as claimed.  \end{proof}

Before discussing the impact of the transversality condition on the solvability of the Born equation for $  \epsilon_r=0$, we need a  result regarding the smoothness of the solution to scattering of transverse vector fields.

\begin{proposition}[Smooth solution to electromagnetic scattering]\label{prop:smoothsln}If the incident field is a smooth and divergence-free vector field defined in a bounded and open volume $V\Subset\mathbb R^3 $, i.e.~$\bm E_\mathrm{inc}\in T(V;\mathbb C^3)= L^2(V;\mathbb C^3)\cap  C^\infty(V;\mathbb C^3) \cap\ker(\nabla\cdot)  $, and $ \bm E\in \Phi(V;\mathbb C^3)=\Cl(T(V;\mathbb C^3))$ is a solution to the Born equation $\bm E_\mathrm{inc}=(\hat I -\chi\hat{\mathscr G}) \bm E$, then the vector field $ \bm E(\bm r),\bm r\in  V$ is also smooth and divergence-free: $\bm E\in T(V;\mathbb C^3)  $.

Additionally, if $ \nabla\times\bm E_{\mathrm{inc}}\in  L^2(V;\mathbb C^3)$, then the same property applies to any solution of the Born equation  $\bm E_\mathrm{inc}=(\hat I -\chi\hat{\mathscr G}) \bm E\in T(V;\mathbb C^3)$:\begin{align}
\iiint_V|\nabla\times\bm E(\bm r)|^2\D^3\bm r<+\infty.\label{eq:fin_mag_E}
\end{align}\end{proposition}
\begin{proof} Before confirming the smoothness of the solution $\bm E\in T(V;\mathbb C^3)  $, we   show that   $(\hat{\mathscr G}-\hat\gamma)\bm E$  is always a smooth vector field  satisfying the Laplace equation, for all  $\bm E\in\Phi(V;\mathbb C^3)$, in the two paragraphs below.

First, we point out that for any  $ \bm E\in T(V;\mathbb C^3)= L^2(V;\mathbb C^3)\cap  C^\infty(V;\mathbb C^3) \cap\ker(\nabla\cdot) $, we have
\begin{align}((\hat{\mathscr G}-\hat\gamma)\bm E)(\bm r)={}&\nabla\times\nabla\times\iiint_V\frac{\bm E(\bm r')}{4\pi|\bm r-\bm r'|}\D^3\bm r'-\bm E(\bm r')\notag\\={}&\nabla\left[ \nabla\cdot\iiint_V\frac{\bm E(\bm r')}{4\pi|\bm r-\bm r'|}\D^3\bm r' \right],\quad\bm r\in V, \end{align}and\begin{align}(\hat{\mathscr G}-\hat\gamma)\bm E \in\mathfrak b^2(V;\mathbb C^3)\colonequals {}&L^2(V;\mathbb C^3)\cap C^\infty(V;\mathbb C^3)\cap\ker(\nabla^2)\notag\\={}&\{\bm F\in L^2(V;\mathbb C^3)\cap C^\infty(V;\mathbb C^3)|\nabla^2\bm F(\bm r)=\mathbf{0},\forall \bm r\in V \}.\end{align} Here,  $\mathfrak b^p(V;\mathbb C^3) $ denotes the Bergman space of power $p\geq1$ \cite[p.~171]{Axler}. To verify the claim above, we pick  an enclosed open ball $O(\bm r_{*},\varepsilon)\subset V $, and apply the Gau{\ss} theorem in vector calculus, as follows:\begin{align}((\hat{\mathscr G}-\hat \gamma)\bm E)(\bm r)={}&\nabla\left[\nabla\cdot\iiint_{V\smallsetminus O(\bm r_{*},\varepsilon) }\frac{\bm E(\bm r')}{4\pi|\bm r-\bm r'|}\D^3\bm r'\right]\notag\\{}&+\varoiint_{\partial O(\bm r_{*},\varepsilon)}[\bm n'\cdot\bm E(\bm r')]\nabla'\frac{1}{4\pi|\bm r-\bm r'|} \D S'.\end{align}For $\bm r\in O(\bm r_{*},\varepsilon/2)$, one can always differentiate under the integral sign to arbitrarily high order, so as  to conclude that  $(\hat{\mathscr G}-\hat\gamma)\bm E\in C^\infty(O(\bm r_{*},\varepsilon/2);\mathbb C^3) $, and the Laplace equation $\nabla^2((\hat{\mathscr G}-\hat\gamma)\bm E)(\bm r)=\mathbf{0},\forall \bm r\in O(\bm r_{*},\varepsilon/2) $ holds. As the choice of  $O(\bm r_{*},\varepsilon) $ is arbitrary, this shows that $\nabla^2((\hat{\mathscr G}-\hat \gamma)\bm E)(\bm r)=\mathbf{0},\forall \bm r\in V $.

Then, we point out that for all  $\bm E\in\Phi(V;\mathbb C^3)=\Cl( T(V;\mathbb C^3))$, we have  $ (\hat{\mathscr G}-\hat\gamma)\bm E \in(\hat{\mathscr G}-\hat\gamma)\Cl(T(V;\mathbb C^3))\subset\Cl((\hat{\mathscr G}-\hat\gamma)T(V;\mathbb C^3))\subset\Cl(\mathfrak b^2(V;\mathbb C^3))=\mathfrak b^2(V;\mathbb C^3) =L^2(V;\mathbb C^3)\cap  C^\infty(V;\mathbb C^3) \cap\ker(\nabla^2)$. This is because $(\hat{\mathscr G}-\hat \gamma):L^2(V;\mathbb C^3)\longrightarrow L^2(V;\mathbb C^3)$ is continuous, and the Bergman space $\mathfrak b^2(V;\mathbb C^3)=\Cl(\mathfrak b^2(V;\mathbb C^3))  $ is a closed subspace of the Hilbert space $ L^2(V;\mathbb C^3)$  \cite[Proposition 8.3]{Axler}. Thus, we have the Born equation\begin{align}\bm E(\bm r)=\bm E_\mathrm{inc}(\bm r)+\chi(\hat{\gamma}\bm E)(\bm r)+\chi((\hat{\mathscr G}-\hat\gamma)\bm E)(\bm r),\quad\forall\bm r\in V,\label{eq:Born_eq_part}\end{align}where  $\bm E\in\Phi(V;\mathbb C^3)=\Cl( T(V;\mathbb C^3)) $ and $(\hat{\mathscr G}-\hat\gamma)\bm E \in\mathfrak b^2(V;\mathbb C^3) $. Using the Cauchy--Schwarz inequality, one can verify that $\hat{\gamma}\bm E\in C^0(\mathbb R^3;\mathbb C^3) $ for all square-integrable vector fields $\bm E\in L^2(V;\mathbb C^3) $. Therefore, all the three terms on the right-hand side of the Born equation belongs to $C^0(V;\mathbb C^3) $, hence $\bm E\in C^0(V;\mathbb C^3) $. Using differentiation under the integral sign, one can check that $\bm E\in C^0(V;\mathbb C^3) \cap L^2(V;\mathbb C^3)  $ entails $\hat{\gamma}\bm E\in C^1(V;\mathbb C^3) $, so  $\bm E=(\hat I -\chi\hat{\mathscr G})^{-1} \bm E_\mathrm{inc}\in C^1(V;\mathbb C^3)$. By induction, we may deduce   $\bm E=(\hat I -\chi\hat{\mathscr G})^{-1} \bm E_\mathrm{inc}\in C^{m+1}(V;\mathbb C^3)$ from the assumption that    $\bm E\in C^{m}(V;\mathbb C^3)$ for all $m=0,1,2,\dots$, using differentiation under the integral sign and integration by parts. Hence, we may conclude that the solution to the Born equation is a smooth vector field $\bm E=(\hat I -\chi\hat{\mathscr G})^{-1} \bm E_\mathrm{inc}\in  L^2(V;\mathbb C^3)\cap  C^\infty(V;\mathbb C^3) $.

 Taking divergence on both sides of the Born equation, one thus confirms that $\bm E\in \ker(\nabla\cdot) $ as well. This shows that the solution   $\bm E\in L^2(V;\mathbb C^3)\cap  C^\infty(V;\mathbb C^3) \cap\ker(\nabla\cdot)=T(V;\mathbb C^3) $ must also be a transverse field.

 Now, suppose that we know $ \bm E_{\mathrm{inc}}\in T(V;\mathbb C^3)$ and $ \nabla\times\bm E_{\mathrm{inc}}\in L^2(V;\mathbb C^3)$. The smoothness of $ \bm E\in C^\infty(V;\mathbb C^3)$ allows us to carry out the following differentiations (as classical derivatives rather than distributional derivatives):{\small\begin{align}&
(1+\chi)\nabla\times\bm E(\bm r)\notag\\={}&\nabla\times\bm E_{\mathrm{inc}}(\bm r)+\chi\nabla\times\nabla\times\nabla\times\iiint_V\frac{\bm E(\bm r') e^{-ik|\bm r-\bm r' |}}{4\pi|\bm r-\bm r' |}\D^3 \bm r'\notag\\={}&\nabla\times\bm E_{\mathrm{inc}}(\bm r)+\chi\nabla\left[\nabla\cdot\nabla\times\iiint_V\frac{\bm E(\bm r') e^{-ik|\bm r-\bm r' |}}{4\pi|\bm r-\bm r' |}\D^3 \bm r'\right]-\chi\nabla^2\left[\nabla\times\iiint_V\frac{\bm E(\bm r') e^{-ik|\bm r-\bm r' |}}{4\pi|\bm r-\bm r' |}\D^3 \bm r'\right]\notag\\={}&\nabla\times\bm E_{\mathrm{inc}}(\bm r)+\chi\nabla\times\bm E(\bm r)+\chi k^{2}\nabla\times\iiint_V\frac{\bm E(\bm r') e^{-ik|\bm r-\bm r' |}}{4\pi|\bm r-\bm r' |}\D^3 \bm r',\quad \bm r\in V.
\end{align}}Now, it is clear that $ \nabla\times\bm E(\bm r)-\nabla\times\bm E_{\mathrm{inc}}(\bm r)$ is equal to\begin{align}
\chi k^{2}\nabla\times\iiint_V\frac{\bm E(\bm r')}{4\pi|\bm r-\bm r' |}\D^3 \bm r'+\chi k^{2}\nabla\times\iiint_V\frac{\bm E(\bm r') (e^{-ik|\bm r-\bm r' |}-1)}{4\pi|\bm r-\bm r' |}\D^3 \bm r'.
\end{align}   The first summand of the equation above is a member of $ L^2(V;\mathbb C^3)$, by analysis of the Newton potential \eqref{eq:Newton_potential}, while the second term also belongs to  $ L^2(V;\mathbb C^3)$, owing to a Hilbert--Schmidt bound similar to that of $ \hat\gamma$. This proves that $ \nabla\times\bm E-\nabla\times\bm E_{\mathrm{inc}}\in L^2(V;\mathbb C^3)$, as claimed.      \end{proof}

The transversality condition makes a striking difference when it comes to the solvability issue for $ \chi=-1$ (\textit{i.e.}~$ \epsilon_r=0$).
 In  Proposition~\ref{prop:sigma01}, we showed that $-1\in\sigma_p(\hat{\mathscr G})$ is an eigenvalue with an infinite-dimensional eigenspace  $\ker(\hat{\mathscr G}+\hat I)$ in $L^2(V;\mathbb C^3)$. We now set out to give complete characterization of the eigenspace $\ker(\hat{\mathscr G}+\hat I) $ (in Proposition~\ref{prop:kerspace}) before we show that  $\ker(\hat{\mathscr G}+\hat I)$ is orthogonal to $\Phi(V;\mathbb C^3) $ under certain geometric and topological constraints (in Proposition~\ref{prop:smooth_bd}), hence $ -1\notin\sigma_p^\Phi(\hat{\mathscr G}) $.

\begin{proposition}[Characterization of eigenspace $ \ker(\hat{\mathscr G}+\hat I)$]\label{prop:kerspace}So long as the volume $V$ is open and bounded, the exterior volume  $\mathbb R^3\smallsetminus(V\cup\partial V)$ is connected and  $\iiint_{\partial V}\D^3\bm r=0 $,   we  have the identity
\begin{align}\ker(\hat{\mathscr G}+\hat I) =\{\bm E\in L^2(V;\mathbb C^3)|\bm q\times\widetilde{\bm E}(\bm q)=\mathbf 0,\forall\bm q\in \mathbb R^3\}.\label{eq:ker_G+I_qxEq} \end{align}
\end{proposition}
\begin{proof}First, we note that so long as the volume $V\subset \mathbb R^3$ is open and bounded, the set inclusion relation $ \ker(\hat{\mathscr G}+\hat I) \supset\{\bm E\in L^2(V;\mathbb C^3)|\bm q\times\widetilde{\bm E}(\bm q)=\mathbf 0,\forall\bm q\in \mathbb R^3\}$ is an immediate consequence of the Fourier inversion formula
\begin{align}((\hat{\mathscr G}+\hat I)\bm E)(\bm r)\colonequals -\frac1{(2\pi)^3}\lim_{\varepsilon\to0^+}\Fint_{\mathbb R^3}\frac{\bm q\times[\bm q\times\widetilde{\bm E}(\bm q)]e^{-i\bm q\cdot\bm r}}{|\bm q|^2-(k-i\varepsilon)^2}\D^3\bm q.\end{align}

Then, we can see that if   $\mathbb R^3\smallsetminus(V\cup\partial V)$ is connected,  the boundary has zero volume  $\iiint_{\partial V}\D^3\bm r=0 $, and $\Vert(\hat{\mathscr G}+\hat I)\bm E\Vert_{L^2(V;\mathbb C^3)}=0$, then the proof Corollary~\ref{thm:uniq}(B)  [in particular, \eqref{eq:qxE_0}] may be adapted to the $\chi=-1$ case, thus  giving rise to
 \begin{align}|\bm q|^2\frac{\bm q\times\widetilde{\bm E}(\bm q)}{|\bm q|^2-k^{2}}=\mathbf 0,\quad\forall\bm q\in \mathbb R^3\smallsetminus V_0'',\quad\text{where }\iiint_{V_0''}\D^3\bm q=0.\end{align}As $\widetilde {\bm E }\in C^\infty(\mathbb R^3;\mathbb C^3)$ is continuous in $\bm q$, we may extend the equality $\bm q\times\widetilde{\bm E}(\bm q)=\mathbf 0$ to all $\bm q\in \mathbb R^3 $. This proves the reverse set inclusion relation  $ \ker(\hat{\mathscr G}+\hat I) \subset\{\bm E\in L^2(V;\mathbb C^3)|\bm q\times\widetilde{\bm E}(\bm q)=\mathbf 0,\forall\bm q\in \mathbb R^3\}$.
 \end{proof}

From now on, we shall assume  that our dielectric volume $V$ and its boundary $ \partial V$ satisfy the topological and geometric requirements
\ref{itm:bd_open_V}--\ref{itm:disjoint_bdy} listed in the opening of \S\ref{sec:div0}. We shall simply refer to such a scenario as  ``smooth dielectric'' for short. \begin{proposition}[R\^ole of $ \chi=-1$ in the physical spectrum: $-1\notin \sigma_p^\Phi(\hat{\mathscr G})$]\label{prop:smooth_bd}For a smooth dielectric with connected  exterior volume   $\mathbb R^3\smallsetminus(V\cup\partial V)$, we have $
\ker(\hat{\mathscr G}+\hat I) \cap\Phi(V;\mathbb C^3)=\{\mathbf 0\}$.
\end{proposition}
\begin{proof}Pick any $ \bm E\in \ker(\hat{\mathscr G}+\hat I) \cap\Phi(V;\mathbb C^3)$, then we have $ \bm 0=(\hat I+\hat{\mathscr G})\bm E$, and $ \bm E\in T(V;\mathbb C^3),\nabla\times\bm E\in L^2(V;\mathbb C^3)$ according to Proposition~\ref{prop:smoothsln}.

Using the tubular neighborhood lemma (see \cite[pp.~109--114 and p.~212]{doCarmo} or \cite[pp.~109--118]{Hirsch:DiffTop}), one can show that the dielectric boundary occupies zero volume $ \iiint_{\partial V}\D^3\bm r=0$; furthermore, there exists a positive number $ \delta_{V}>0$, such that for all  $ \varepsilon\in(0,\delta_{V})$,  the ``pinched volume'' $ V^-_{(\varepsilon)}\colonequals V\smallsetminus\{\bm r-\varepsilon'\bm n(\bm r)|\bm r\in\partial V,0\leq \varepsilon'\leq\varepsilon \}$ is homeomorphic to $ V$,  and the boundary $ \partial V^{-}_{(\varepsilon)}$ of the ``pinched volume''  is a smooth 2-manifold. (The same is true when we replace $ V^-_{(\varepsilon)}$ and its boundary $ \partial V^{-}_{(\varepsilon)}$  by $ V^+_{(\varepsilon)}$ and $ \partial V^+_{(\varepsilon)}$, where $V^+_{(\varepsilon)}\colonequals V\cup\{\bm r+\varepsilon'\bm n(\bm r)|\bm r\in\partial V,0\leq \varepsilon'<\varepsilon \}$.)

On one hand, for any  $ \varepsilon\in(0,\delta_{V})$, we can use vector calculus to deduce the following relation for all $ \bm q\in\mathbb R^3$:\begin{align}
\varoiint_{\partial V^{-}_{(\varepsilon)}}\bm n\times [\bm E(\bm r)e^{i\bm q\cdot \bm r}]\D S={}&\iiint_{ V^{-}_{(\varepsilon)}}\nabla\times[\bm E(\bm r)e^{i\bm q\cdot \bm r}]\D^3\bm r\notag\\={}&\iiint_{V^{-}_{(\varepsilon)}}e^{i\bm q\cdot \bm r}\nabla\times \bm E(\bm r)\D^3\bm r+i\bm q\times\iiint_{V^{-}_{(\varepsilon)}}e^{i\bm q\cdot \bm r}\bm E(\bm r)\D^3\bm r.
\end{align} As $ \varepsilon\to0^+$, the last term in the equation above tends to $ i\bm q\times\widetilde{\bm E}(\bm q) =\bm 0$, according to \eqref{eq:ker_G+I_qxEq}.  Meanwhile, since $ \nabla\times\bm E\in L^2(V;\mathbb C^3)$, we know that the following limit exists:\begin{align}
\lim_{\varepsilon\to0^+}\varoiint_{\partial V^{-}_{(\varepsilon)}}e^{i\bm q\cdot \bm r}\bm n\times \bm E(\bm r)\D S=\iiint_{V}e^{i\bm q\cdot \bm r}\nabla\times \bm E(\bm r)\D^3\bm r.\label{eq:dV_V_FT}
\end{align}Here, the left-hand side of \eqref{eq:dV_V_FT} can be rewritten as $ \varoiint_{\partial V}e^{i\bm q\cdot \bm r}\bm n\times \bm E(\bm r)\D S$, with the understanding that the surface integral is in fact the canonical pairing between $ e^{i\bm q\cdot \bm r}\in H^{1/2}(\partial V;\mathbb C)$ and $\bm n\times \bm E(\bm r)\in H^{-1/2}(\partial V;\mathbb C^3)$ [cf.~\eqref{eq:bd_tr_can_pair}]. Multiplying both sides of  \eqref{eq:dV_V_FT} by $|\bm q|^{-2} e^{-i\bm q\cdot\bm r'-|\bm q|\kappa}$ for $ \bm r'\in V,\kappa>0$ and integrating over $ \bm q\in\mathbb R^3$, we obtain the following result in the $ \kappa\to0^+$ limit:\begin{align}
\varoiint_{\partial V}\frac{\bm n\times \bm E(\bm r)}{4\pi|\bm r-\bm r'|}\D S=\iiint_{V}\frac{\nabla\times \bm E(\bm r)}{4\pi|\bm r-\bm r'|}\D^3\bm r,\quad \bm r'\in V.
\end{align}Hitting the Laplace operator on both sides, we obtain $ \nabla\times\bm E(\bm r)=\bm 0,\forall\bm r\in V$. Subsequently, a back substitution in \eqref{eq:dV_V_FT} leads us to  \begin{align}
\varoiint_{\partial V}\frac{\bm n'\times \bm E(\bm r')}{4\pi|\widetilde{\bm r}-\bm r'|}\D S'=\bm 0,\quad\forall\widetilde{\bm r}\in \mathbb R^3\smallsetminus\partial V.
\end{align} If we approach the dielectric boundary $ \partial V$ from the inside, setting  $\widetilde{ \bm r}=\bm r-\varepsilon\bm n,\varepsilon\to0^+$ for $ \bm r\in\partial V$ in the equation above, then we have\begin{subequations} \begin{align}
\mathbf 0=&\lim_{\varepsilon\to0^+}(\bm n\cdot\nabla)\varoiint_{\partial V}\bm n'\times \bm E(\bm r'){\frac{1}{4\pi|\bm r-\varepsilon\bm n-\bm r'|}}\D S'\notag\\\colonequals{}&  -\lim_{\varepsilon\to0^+}\frac{\partial}{\partial \varepsilon}\varoiint_{\partial V}\bm n'\times \bm E(\bm r'){\frac{1}{4\pi|\bm r-\varepsilon\bm n-\bm r'|}}\D S'\notag\\={}&\varoiint_{\partial V}\bm n'\times \bm E(\bm r'){(\bm n\cdot\nabla)\frac{1}{4\pi|\bm r-\bm r'|}}\D S'+\frac{1}{2}\bm n\times \bm E(\bm r),\quad\bm r\in \partial V,\label{eq:Et_-}
\intertext{where the limit is interpreted as a boundary trace. If we repeat this procedure
with   $\widetilde{ \bm r}=\bm r+\varepsilon\bm n,\varepsilon\to0^+$, then we have}
\mathbf 0=&\lim_{\varepsilon\to0^+}(\bm n\cdot\nabla)\varoiint_{\partial V}\bm n'\times \bm E(\bm r'){\frac{1}{4\pi|\bm r+\varepsilon\bm n-\bm r'|}}\D S'\notag\\\colonequals{}& +\lim_{\varepsilon\to0^+}\frac{\partial}{\partial \varepsilon}\varoiint_{\partial V}\bm n'\times \bm E(\bm r'){\frac{1}{4\pi|\bm r+\varepsilon\bm n-\bm r'|}}\D S'\notag\\={}&\varoiint_{\partial V}\bm n'\times \bm E(\bm r'){(\bm n\cdot\nabla)\frac{1}{4\pi|\bm r-\bm r'|}}\D S'-\frac{1}{2}\bm n\times \bm E(\bm r),\quad\bm r\in \partial V.\label{eq:Et_+}
\end{align}\end{subequations}Comparing \eqref{eq:Et_-} and \eqref{eq:Et_+}, we see that the boundary trace $\bm n\times \bm E(\bm r),\bm r\in \partial V $ vanishes.

On the other hand, the transverse  field in question $ \bm E\in T(V;\mathbb C^3)$ satisfies the vector Laplace equation $ \nabla^2\bm E(\bm r)=\nabla[\nabla\cdot\bm E(\bm r)]-\nabla\times[\nabla\times\bm E(\bm r)]=\bm 0,\bm r\in V$, so its Green identity
\begin{align}\bm E(\bm r)=\varoiint_{\partial  V^{-}_{(\varepsilon)}}\left[ \frac{1}{4\pi|\bm r-\bm r'|}(\bm n'\cdot\nabla')\bm E(\bm r')-\bm E(\bm r')(\bm n'\cdot\nabla')\frac{1}{4\pi|\bm r-\bm r'|} \right]\D S',\quad\forall\bm r\in  V^{-}_{(\varepsilon)}\end{align}can be modified into the following ``Kirchhoff diffraction integral'' (see the argument in \cite[p.~483]{Jackson:EM}):
{\small\begin{align}&\bm E(\bm r)\notag\\={}&-\varoiint_{\partial  V^{-}_{(\varepsilon)}}\left\{ \frac{\bm n'\times[\nabla'\times\bm E(\bm r')]}{4\pi|\bm r-\bm r'|}+[\bm n'\times\bm E(\bm r')]\times\nabla'\frac{1}{4\pi|\bm r-\bm r'|}+[\bm n'\cdot\bm E(\bm r')]\nabla'\frac{1}{4\pi|\bm r-\bm r'|} \right\}\D S'\notag\\={}&-\varoiint_{\partial  V^{-}_{(\varepsilon)}}\left\{[\bm n'\times\bm E(\bm r')]\times\nabla'\frac{1}{4\pi|\bm r-\bm r'|}+[\bm n'\cdot\bm E(\bm r')]\nabla'\frac{1}{4\pi|\bm r-\bm r'|}\right\}\D S',\quad\forall\bm r\in  V^{-}_{(\varepsilon)},\end{align}}where $\bm n$ is the outward unit normal vector on the boundary $\partial V^{-}_{(\varepsilon)}$. Sending the positive infinitesimal $\varepsilon$ to zero, and bearing in mind that  $\bm n'\times \bm E(\bm r')=\mathbf 0,\bm r'\in \partial V $, we see that $ \bm E(\bm r)=\nabla u(\bm r),\bm r\in V$, where\begin{align}
u(\bm r)=\varoiint_{\partial V}\frac{\bm n'\cdot\bm E(\bm r')}{4\pi|\bm r-\bm r'|}\D S',\quad\forall\bm r\in V
\end{align}is defined by considering $\bm n'\cdot\bm E(\bm r') \in H^{-1/2}(\partial V;\mathbb C)$ as a boundary trace. Again, by  $\mathbf 0=\bm n\times \bm E(\bm r)=\bm n\times\nabla u(\bm r),\bm r\in \partial V $, we know that $  u(\bm r)$ remains constant on each connected component of $ \partial V$.

Now that   the exterior volume   $\mathbb R^3\smallsetminus(V\cup\partial V)$ is connected,  the boundary $ \partial V_j$ for each connected component $ V_j$ of the dielectric volume $V=\bigcup_{j=1}^{N_V}V_j$ must also be connected, according to the Alexander relation \cite[pp.~384--387]{DautrayLionsVol3}. Therefore, we can put down      $ u(\bm r)=u_j,\bm r\in\partial V_j,j\in\{1,\dots,N_V\}$ with constants $ u_j,j\in\{1,\dots,N_V\}$. For each fixed $j$, we have \begin{align}
0={}&\varoiint_{\partial V_j}[u(\bm r)- u_j](\bm n\cdot\nabla)u^{*}(\bm r)\D S=\iiint_{V_j}\{[u(\bm r)- u_j]\nabla^2u^{*}(\bm r)+|\nabla u(\bm r)|^{2}\}\D^3\bm r\notag\\={}&\iiint_{V_j}|\nabla u(\bm r)|^{2}\D^3\bm r=\iiint_{V_j}|\bm E(\bm r)|^{2}\D^3\bm r.
\end{align}This proves $ \bm E(\bm r)=\mathbf 0,\bm r\in V$, as claimed.
 \end{proof}

\begin{remark}
The foregoing proof can be readily generalized to the following result:\begin{align}
\dim[\ker(\hat{\mathscr G}+\hat I)\cap\Phi(V;\mathbb C^3)]={}&\dim\mathscr H^0(\partial V)-\dim\mathscr H^0( V\cup \partial V)\notag\\={}&\dim\mathscr H^0(\mathbb R^3\smallsetminus V)-1,
\end{align}where $ \dim\mathscr H^0(\partial V)-\dim\mathscr H^0( V\cup \partial V)$ counts the difference between the numbers of path-connected components of $ \partial V$ and $ V\cup\partial V$, and the second equality is just a statement of the Alexander relation \cite[pp.~384--387]{DautrayLionsVol3}.
\eor
\end{remark}

In short, the uniqueness theorem $ -1\notin \sigma_p^\Phi(\hat{\mathscr G})$  depends on some subtle geometric and topological properties of smooth surfaces.

\subsection{Polynomial compactness and optical resonance\label{subsec:cpt_poly}}
  Now we will state and prove the ``optical resonance theorem'' --- a  result  that  outlines the structure of the physical spectrum $ \sigma^\Phi(\hat{\mathscr G})$.  \begin{theorem}[Optical resonance  in electromagnetic scattering]\label{thm:optres}For a smooth  dielectric volume $V\Subset\mathbb R^3 $,   the   operator $\hat {\mathscr G}(\hat I+2\hat{\mathscr G})\colon \Phi(V;\mathbb C^3)\longrightarrow \Phi(V;\mathbb C^3) $ is compact, and $\sigma^\Phi(\hat{\mathscr G})\subset \sigma^\Phi_p(\hat{\mathscr G})\cup\{0,-1/2\} $.
  \end{theorem}
  \begin{proof}  Using the technique of boundary traces, we will show  that the   operator $\hat {\mathscr G}(\hat I+2\hat{\mathscr G})\colon\Phi(V;\mathbb C^3)\longrightarrow \Phi(V;\mathbb C^3) $ is indeed compact.
What remains then follows from the spectral mapping theorem \cite[p.~227]{Yosida} and the Riesz--Schauder theory for the spectrum of compact operators  \cite[pp.~283--284]{Yosida}.

 We note that
  \begin{align}\hat {\mathscr G}(\hat I+2\hat{\mathscr G})-(\hat {\mathscr G}-\hat \gamma)(\hat I+2(\hat {\mathscr G}-\hat \gamma))=\hat \gamma+2\hat {\mathscr G}\hat \gamma+2\hat\gamma\hat {\mathscr G}-2\hat \gamma^2\label{eq:cptdiff1}\end{align} is  a compact operator, as each addend on the right-hand side of \eqref{eq:cptdiff1} is a composition of a compact operator with a bounded operator, thus also a compact operator in its own right.
    Therefore, we may establish the compactness of the   operator $\hat {\mathscr G}(\hat I+2\hat{\mathscr G}):\Phi(V;\mathbb C^3)\longrightarrow \Phi(V;\mathbb C^3) $ by proving that the operator  $(\hat {\mathscr G}-\hat \gamma)(\hat I+2(\hat {\mathscr G}-\hat \gamma)):\Phi(V;\mathbb C^3)\longrightarrow \Phi(V;\mathbb C^3) $ is compact.

A key observation is that, the operator $ \hat{\mathscr G}-\hat\gamma$ maps any $ \bm E\in\Phi(V;\mathbb C^3) $ to \begin{align}((\hat{\mathscr G}-\hat\gamma)\bm E)(\bm r)=-\nabla\left[\varoiint_{\partial V}\frac{{\bm n'\cdot\bm E(\bm r')} }{4\pi|\bm r-\bm r' |}\D S'\right],\end{align}which is the gradient of a harmonic function.  The bulk behavior $((\hat{\mathscr G}-\hat\gamma)\bm E)(\bm r),\bm r\in V$  is fully determined by the boundary behavior of its normal component  $\bm n\cdot((\hat{\mathscr G}-\hat\gamma)\bm E)(\bm r),\bm r\in \partial V$, and this one-to-one correspondence is also robust in both directions  \cite[p.~252]{DautrayLionsVol3}. Technically speaking, according to the boundary trace theorem and the robustness of Neumann boundary problems, there are two finite positive constants $ C_1$ and $C_2$ such that\begin{align}C_1\Vert(\hat{\mathscr G}-\hat\gamma)\bm E\Vert_{L^2(V;\mathbb C^3)}\leq \Vert\bm n\cdot((\hat{\mathscr G}-\hat\gamma)\bm E)\Vert_{H^{-1/2}(\partial V;\mathbb C)}\leq C_2\Vert(\hat{\mathscr G}-\hat\gamma)\bm E\Vert_{L^2(V;\mathbb C^3)}.\end{align}

In view of the commutative diagram  \begin{align}\begin{split}
{\xymatrix{\bm E\in L^2(V;\mathbb C^3)\ar@{->}[d]\\(\hat{\mathscr G}-\hat\gamma)\bm E\in L^2(V;\mathbb C^3)\ar@<.5ex>[d]\ar@{->}[r]&(\hat{\mathscr G}-\hat\gamma)\bm E+2((\hat{\mathscr G}-\hat\gamma)^{2}\bm E)\in L^2(V;\mathbb C^3)\ar@<.5ex>[d]\\\bm n\cdot((\hat{\mathscr G}-\hat\gamma)\bm E)\in H^{-1/2}(\partial V;\mathbb C)\ar@<.5ex>[u]\ar@{->}[r]^-{(*)}&\bm n\cdot((\hat{\mathscr G}-\hat\gamma)\bm E)+2\bm n\cdot((\hat{\mathscr G}-\hat\gamma)^{2}\bm E)\in H^{-1/2}(\partial V;\mathbb C)\ar@<.5ex>[u]}}\end{split}
\end{align}(where the vertical arrows between the last two rows are homeomorphisms connecting bulk and boundary behavior for gradients of harmonic functions), the compactness of the  polynomial $(\hat{\mathscr G}-\hat \gamma)+2(\hat{\mathscr G}-\hat\gamma)^2 $ on   $\Phi(V;\mathbb C^3) $ descends from the compactness of the mapping marked with  ($*$), the latter of which is evident from the boundary integral representation\begin{align}&\bm n\cdot((\hat{\mathscr G}-\hat\gamma)\bm E)(\bm r)+2\bm n\cdot((\hat{\mathscr G}-\hat\gamma)^{2}\bm E)(\bm r)\notag\\={}&-2\varoiint_{\partial V}\bm n'\cdot((\hat{\mathscr G}-\hat\gamma)\bm E)(\bm r'){(\bm n\cdot\nabla)\frac{1}{4\pi|\bm r-\bm r'|}}\D S',\quad\bm r\in \partial V\label{eq:dV_int}\end{align}where the integral kernel $(\bm n\cdot\nabla)(4\pi|\bm r-\bm r'|)^{-1}=O(|\bm r-\bm r'|^{-1}) $ is weakly singular (see \cite[p.~48]{ColtonKress1983}  or \cite{HsiaoKleinman}), hence induces a compact operator (cf.~\cite{HsiaoKleinman} or \cite[p.~124]{DautrayLionsVol4}) on $ H^{-1/2}(\partial V;\mathbb C)$.

Here, the choice of the coefficient $2$ in the quadratic term of $(\hat{\mathscr G}-\hat \gamma)+2(\hat{\mathscr G}-\hat\gamma)^2 $  is critical, in that \begin{align}&\lim_{\varepsilon\to0^+}(\bm n\cdot\nabla)\varoiint_{\partial V}\bm n'\cdot((\hat{\mathscr G}-\hat\gamma)\bm E)(\bm r'){\frac{1}{4\pi|\bm r-\varepsilon\bm n-\bm r'|}}\D S'\notag\\={}&\varoiint_{\partial V}\bm n'\cdot((\hat{\mathscr G}-\hat\gamma)\bm E)(\bm r'){(\bm n\cdot\nabla)\frac{1}{4\pi|\bm r-\bm r'|}}\D S'+\frac{1}{2}\bm n\cdot((\hat{\mathscr G}-\hat\gamma)\bm E)(\bm r),\quad\bm r\in \partial V.\label{eq:dV_1/2}\end{align}Here, as in the proof of Proposition \ref{prop:smooth_bd}, the limit is interpreted in the sense of boundary trace.
 \end{proof}\begin{remark}
In \cite[Chapter 4 and Appendix C]{ZhouThesis}, we originally proved  the theorem above by working directly with the operator   $(\hat {\mathscr G}-\hat \gamma)(\hat I+2(\hat {\mathscr G}-\hat \gamma)):\Phi(V;\mathbb C^3)\longrightarrow \Phi(V;\mathbb C^3) $ in the volume integral formulation. After lengthy and arduous arguments, we established the compactness of such a volume integral operator via uniform approximation of compact operators and verification of the Calder\'on--Zygmund cancellation condition. In the short proof presented here using the boundary integral operators, the arguments are made more transparent: the construction of the quadratic polynomial ensures that the non-compact ingredients [constant multiples of the identity map on the infinite-dimensional space $ H^{-1/2}(\partial V;\mathbb C)$] cancel out.\eor\end{remark}\begin{remark}\label{rmk:cpt_G_miscncpt}We note that our compactness argument for the boundary traces on the fractional order Sobolev space $ H^{-1/2}(\partial V;\mathbb C)$ follows the ideas of Hsiao--Kleinman \cite[Sect.~VI]{HsiaoKleinman} and N\'ed\'elec \cite[\S5.6]{Nedelec2001}. Such a compactness argument also played essential r\^oles in the $ H(\curl, V)$ formulation \cite{Ammari2016} of electromagnetic scattering.     \eor\end{remark}

We now point out that any non-vanishing compact polynomial in the Green operator  $ \hat {\mathscr G}:\Phi(V;\mathbb C^3)\longrightarrow \Phi(V;\mathbb C^3)$ must contain the factor  $ \hat {\mathscr G}(\hat I+2\hat{\mathscr G})$.
\begin{proposition}[Minimal compact polynomial of the Green operator]\label{thm:optres1}For a bounded and open  dielectric volume $V\Subset\mathbb R^3 $ with smooth boundary $ \partial V$ and connected exterior volume $ \mathbb R^3\smallsetminus(V\cup\partial V)$,   the monic polynomial $\hat {\mathscr G}(\hat I+2\hat{\mathscr G})/2 $ is the minimal compact polynomial for $ \hat {\mathscr G}\colon \Phi(V;\mathbb C^3)\longrightarrow \Phi(V;\mathbb C^3)$, and $\sigma^\Phi(\hat{\mathscr G})= \sigma^\Phi_p(\hat{\mathscr G})\cup\{0,-1/2\} $.\end{proposition} \begin{proof}To demonstrate the minimality of the polynomial  $\hat {\mathscr G}(\hat I+2\hat{\mathscr G})/2 $, we need to show that any non-zero compact polynomial in $\hat{\mathscr G} $ must contain  $\hat {\mathscr G}(\hat I+2\hat{\mathscr G})/2 $ as a factor. This requires us to prove that  $z\hat I+\hat{\mathscr G} $ is not compact for any $ z\in\mathbb C$.

First, we point out that   $z\hat I+\hat{\mathscr G}\colon \Phi(V;\mathbb C^3)\longrightarrow \Phi(V;\mathbb C^3) $ is not compact for $z\neq0 $. If the converse were true, then     $ \hat{\mathscr G}=z\hat I+\hat{\mathscr G}-z\hat I\colon \Phi(V;\mathbb C^3)\longrightarrow \Phi(V;\mathbb C^3) $ would be a Fredholm operator of index zero, contradicting the fact that $0\in\sigma^\Phi_c(\hat{\mathscr G}) $ (by extension of the proofs for Propositions \ref{prop:sigma01} and \ref{prop:sigma_conn}), namely $ \ker(\hat{\mathscr G})\cap\Phi(V;\mathbb C^3) =\{\mathbf 0\}$, $ \ran(\hat{\mathscr G})\cap\Phi(V;\mathbb C^3) \neq\Phi(V;\mathbb C^3) $, $ \Cl(\ran(\hat{\mathscr G})\cap\Phi(V;\mathbb C^3) )=\Phi(V;\mathbb C^3)$.

If    $ \hat{\mathscr G}\colon \Phi(V;\mathbb C^3)\longrightarrow \Phi(V;\mathbb C^3) $ were compact, then we would be able to construct another compact operator $\hat{\mathscr G}-\hat{\gamma}\colon \Cl((\hat{\mathscr G}-\hat{\gamma})\Phi(V;\mathbb C^3))\longrightarrow \Cl((\hat{\mathscr G}-\hat{\gamma})\Phi(V;\mathbb C^3)) $.  Our proof developed for the compactness of $(\hat{\mathscr G}-\hat\gamma)+2(\hat{\mathscr G}-\hat\gamma) ^2\colon \Phi(V;\mathbb C^3)\longrightarrow\Phi(V;\mathbb C^3) $  can be adapted to show that $ \hat I+2(\hat{\mathscr G}-\hat\gamma)$ is compact on   $ \Cl((\hat{\mathscr G}-\hat{\gamma})\Phi(V;\mathbb C^3))$. Therefore, the mapping    $\hat{\mathscr G}-\hat{\gamma}\colon \Cl((\hat{\mathscr G}-\hat{\gamma})\Phi(V;\mathbb C^3))\longrightarrow \Cl((\hat{\mathscr G}-\hat{\gamma})\Phi(V;\mathbb C^3)) $ is a Fredholm operator on an infinite-dimensional Hilbert space, which cannot be compact. This contradiction shows that neither  $\hat{\mathscr G}-\hat\gamma\colon \Phi(V;\mathbb C^3)\longrightarrow \Phi(V;\mathbb C^3) $ nor   $\hat{\mathscr G}\colon \Phi(V;\mathbb C^3)\longrightarrow \Phi(V;\mathbb C^3) $  is a compact operator.

From the  arguments above, we see that    $(\hat{\mathscr G}-\hat\gamma)^2+\frac12(  \hat{\mathscr G}-\hat\gamma) $ is the minimal compact polynomial for the polynomially compact operator
  $\hat{\mathscr G}-\hat\gamma\colon \Phi(V;\mathbb C^3)\longrightarrow \Phi(V;\mathbb C^3) $. According to the structure theorem of polynomially compact operators \cite{Gilfeather}, we may conclude that any one of the polynomial roots (\textit{i.e.}~$\lambda_{c}=0 ,-1/2$) must fall into one of the following two categories:\begin{enumerate}[label=(\arabic*), widest=a]
\item An isolated point in the point spectrum $\sigma_{p}^\Phi(\hat{\mathscr G}-\hat\gamma) $ with infinite-dimensional eigenspace: $ \dim\ker(\lambda_c\hat I-\hat{\mathscr G}-\hat \gamma)=+\infty$;\item
An accumulation point of the point spectrum $\sigma_{p}^\Phi(\hat{\mathscr G}-\hat\gamma) $.
\end{enumerate} If situation (1) happens, then $\hat I+2(\hat{\mathscr G}-\hat \gamma )$  does not fit the definition of a Fredholm operator; if situation (2) happens, then  $\hat I+2(\hat{\mathscr G}-\hat \gamma )$ does not have closed range, which also disqualifies it as a Fredholm operator. As a consequence, the operator $\hat I+2\hat{\mathscr G} $ is not a Fredholm operator, and we must have $-1/2\in\sigma_c^\Phi(\hat{\mathscr G}) $. This completes the proof that $ \sigma_c^\Phi(\hat{\mathscr G})=\{0,-1/2\}$.
 \end{proof}

In summary, the inevitable singular behavior of the Born equation (\textit{i.e.} ``optical resonance behavior'' of the Maxwell equations) at the critical susceptibility $ \chi=-2\Leftrightarrow \epsilon_r=-1$ has an algebraic origin, and does not depend on the detailed geometric shape of the dielectric.

\subsection{Harmonized Green functions and the Hilbert--Schmidt  polynomial $ \hat {\mathscr G}(\hat I+2\hat {\mathscr G})^2$\label{subsec:HS_poly}}The electrostatic Green function $ G_D(\bm r,\bm r')$ with  Dirichlet boundary condition  is given by the unique solution to
\begin{align}\nabla ^{2}G_D(\bm r,\bm r')=-\delta(\bm r-\bm r'),\quad\bm r,\bm r'\in V;\qquad G_D(\bm r,\bm r')=0,\quad\bm r\in V,\bm r'\in\partial V.\end{align} The Green function thus defined automatically honors reciprocal symmetry  $G_D(\bm r,\bm r')=G_D(\bm r',\bm r)  $  \cite[p.~40]{Jackson:EM}. Customarily, the electrostatic Green function  $G_N(\bm r,\bm r') $ with Neumann boundary condition is prescribed as a solution to\begin{align}\nabla ^{2}G_N(\bm r,\bm r')=-\delta(\bm r-\bm r'),\;\;\bm r,\bm r'\in V;\quad (\bm n'\cdot\nabla')G_N(\bm r,\bm r')=-\frac{1}{\varoiint_{\partial V}\D S'},\;\;\bm r\in V,\bm r'\in\partial V.\end{align}Hereafter, we will impose the constraint of reciprocal symmetry to the Green function $G_N(\bm r,\bm r')=G_N(\bm r',\bm r) $ (see \cite{KimJackson1993} or \cite[p.~40]{Jackson:EM} for such feasibility).

In the next short lemma, we will derive a volume integral analog of \eqref{eq:dV_int},  based on  electrostatic Green functions. \begin{lemma}[Harmonized electrostatic Green function] Define\begin{align}\mathfrak g(\bm r,\bm r')\colonequals G_D(\bm r,\bm r')+G_N(\bm r,\bm r')-\frac{1}{2\pi|\bm r-\bm r'|},\quad \bm r,\bm r'\in V,\end{align}which is a harmonic function with respect to both $\bm r$ and $\bm r'$, then we have  the volume integral representation\begin{align}((\hat{\mathscr G}-\hat\gamma)\bm E)(\bm r)+2((\hat{\mathscr G}-\hat\gamma)^{2}\bm E)(\bm r)=\iiint_V\nabla\nabla'\mathfrak g(\bm r,\bm r')\cdot((\hat{\mathscr G}-\hat\gamma)\bm E)(\bm r')\D^3\bm r',\quad\bm r\in V,\label{eq:GN_1/2}\end{align}where  $ \nabla\nabla'\mathfrak g(\bm r,\bm r')$ is  a $ 3\times3$ matrix filled with mixed second-order derivatives of $\mathfrak g(\bm r,\bm r') $. \end{lemma}\begin{proof}By properties of the Green functions, we have\begin{align}&\iiint_V\nabla\nabla'\mathfrak g(\bm r,\bm r')\cdot((\hat{\mathscr G}-\hat\gamma)\bm E)(\bm r')\D^3\bm r'=\nabla\varoiint_{\partial V}\mathfrak g(\bm r,\bm r')\bm n'\cdot((\hat{\mathscr G}-\hat\gamma)\bm E)(\bm r')\D S'\notag\\={}&\nabla\varoiint_{\partial V}G_N(\bm r,\bm r')\bm n'\cdot((\hat{\mathscr G}-\hat\gamma)\bm E)(\bm r')\D S'-2\nabla\varoiint_{\partial V}\frac{\bm n'\cdot((\hat{\mathscr G}-\hat\gamma)\bm E)(\bm r')}{4\pi|\bm r-\bm r'|}\D S'\notag\\={}&((\hat{\mathscr G}-\hat\gamma)\bm E)(\bm r')+2((\hat{\mathscr G}-\hat\gamma)^{2}\bm E)(\bm r),\quad\bm r\in V,\end{align} as claimed in \eqref{eq:GN_1/2}. \end{proof}

From the arguments above, it is straightforward to check that the integral kernel $\hat K_1(\bm r,
\bm r')\colonequals  \nabla\nabla'\mathfrak g(\bm r,\bm r')$ is Hermitian $ \hat K_1^*(\bm r,\bm r')=\hat K_1(\bm r',\bm r)$, where an asterisk denotes the conjugate transpose of a $3\times3$ matrix.
What we have proved in Theorem~\ref{thm:optres} is that  the integral kernel $ \nabla\nabla'\mathfrak g(\bm r,\bm r')$ induces a compact linear operator, unlike the strongly singular integral kernel associated with the non-compact operator $ \hat{\mathscr G}$. In the next section, we will show that the following integral kernel
(where products represent matrix multiplications)\begin{align}\iiint_{V}\nabla\nabla''\mathfrak g(\bm r,\bm r'')\nabla''\nabla'\mathfrak g(\bm r'',\bm r')\D^3\bm r''=\iiint_{V}\hat K_1(\bm r,
\bm r'')\hat K_1(\bm r'',\bm r')\D^3\bm r''\end{align}is square integrable, so it induces a Hilbert--Schmidt operator, just as the ``dynamic correction'' operator $\hat\gamma $.

According to the Schur--Weyl inequality (see \cite{Schur1909,Weyl1949}, also \cite[p.~8]{Simon2005}), we have\begin{align}\sqrt{\vphantom{\sum_{a}^{_a}}\smash{\sum_{\lambda\in\sigma^\Phi(\hat{\mathscr G})}^{\phantom{a}}|\lambda|^2|1+2\lambda|^4}}\leq\Vert\hat{\mathscr G }(\hat I+2\hat{\mathscr G})^2\Vert_2\colonequals \sqrt{\sum_{s=1}^\infty\Vert\hat{\mathscr G }(\hat I+2\hat{\mathscr G})^2\bm e_s\Vert_{L^2(V;\mathbb C^3)}^2}\end{align}where $ \{\bm e_s|s=1,2,\dots\}$ is any complete set of orthonormal basis for the Hilbert space $ \Phi(V;\mathbb C^3)$, so the Hilbert--Schmidt bound $ \Vert\hat{\mathscr G }(\hat I+2\hat{\mathscr G})^2\Vert_2<+\infty$ would entail the convergence of the spectral series in question. Using the inequalities $ \Vert\hat A\hat B\Vert_2\leq\Vert\hat A\Vert_2\Vert\hat B\Vert$ and $\Vert\hat B\hat A\Vert_2\leq\Vert\hat A\Vert_2\Vert\hat B\Vert $ \cite[p.~218]{ReedSimonVolI} (where $ \Vert \hat A\Vert\colonequals\sup_{\bm F\in L^2(V;\mathbb C^3)\smallsetminus\{\mathbf 0\}}\Vert\hat A\bm F\Vert_{L^2(V;\mathbb C^3)}/\Vert\bm F\Vert_{L^2(V;\mathbb C^3)}$), together with the facts that $\Vert\hat{\mathscr G}-\hat\gamma\Vert\leq1 $, $\Vert\hat I+\hat{\mathscr G}-\hat\gamma \Vert\leq1 $ [both from Proposition \ref{prop:Gg_FT}(b)] and $ \Vert\hat\gamma\Vert\leq\Vert\hat\gamma\Vert_2<+\infty$, we may deduce the inequality\begin{align}&\Vert(\hat I+2\hat{\mathscr G})^2\hat{\mathscr G }\Vert_2\notag\\\leq{}&\Vert(\hat I+2\hat{\mathscr G}-2\hat\gamma)^2(\hat{\mathscr G}-\hat\gamma)\Vert_2+\Vert\hat \gamma+4[(\hat{\mathscr G}-\hat\gamma)\hat \gamma+\hat\gamma(\hat{\mathscr G}-\hat\gamma)](\hat I+\hat{\mathscr G}-\hat\gamma)+4(\hat{\mathscr G}-\hat\gamma)^{2}\hat \gamma+\notag\\{}&+4(\hat I+\hat{\mathscr G}-\hat\gamma)\hat \gamma^2+4\hat\gamma(\hat{\mathscr G}-\hat\gamma)\hat\gamma+4\hat \gamma^2(\hat{\mathscr G}-\hat\gamma)+4\hat\gamma^3\Vert_2\notag\\\leq{}&\Vert(\hat I+2\hat{\mathscr G}-2\hat\gamma)^2(\hat{\mathscr G}-\hat\gamma)\Vert_2+\Vert\hat\gamma\Vert_2(13+12\Vert\hat\gamma\Vert+4\Vert\hat\gamma\Vert^2).\label{eq:HS_est}\end{align}Therefore, the major task in Theorem~\ref{thm:HS} boils down to an analysis of the action of $(\hat I+2\hat{\mathscr G}-2\hat\gamma)^2 $ on the range of $ (\hat{\mathscr G}-\hat\gamma)$, which is represented by  the integral kernel\begin{align}\iiint_{V}\nabla\nabla''\mathfrak g(\bm r,\bm r'')\nabla''\nabla'\mathfrak g(\bm r'',\bm r')\D^3\bm r''.\end{align}
Before   establishing the Hilbert--Schmidt bound for the operator $ (\hat I+2\hat{\mathscr G}-2\hat\gamma)^2(\hat{\mathscr G}-\hat\gamma)$ in Proposition~\ref{prop:HS_proof},  we explore  the boundary behavior of the integral kernel  $\mathfrak g(\bm r,\bm r')$ in the lemma below.\begin{lemma}[Boundary behavior of harmonized Green function]\label{lm:bd_g}For any $ f\in H^{-1/2}(\partial V;\mathbb C)$, we have the following limits in the sense of boundary trace:{\small\begin{align}\lim_{\varepsilon\to0^+}\varoiint_{\partial V} f(\bm r')(\bm n'\cdot \nabla') \mathfrak g(\bm r-\varepsilon\bm n,\bm r')\D S'={}&-\varoiint_{\partial V}  f(\bm r')\left[(\bm n'\cdot \nabla')\frac{1}{2\pi|\bm r-\bm r'|}+\frac{1}{\varoiint_{\partial V}\D S}\right]\D S';\label{eq:star}\\\lim_{\varepsilon\to0^+}(\bm n\cdot \nabla)\varoiint_{\partial V}  \mathfrak g(\bm r-\varepsilon\bm n,\bm r')f(\bm r')\D S'={}&-\varoiint_{\partial V}  f(\bm r')\left[(\bm n\cdot \nabla)\frac{1}{2\pi|\bm r-\bm r'|}+\frac{1}{\varoiint_{\partial V}\D S}\right]\D S',\label{eq:2star}\end{align}}where $ \bm r\in \partial V$.\end{lemma}\begin{proof}We first prove \eqref{eq:star} by computing{\allowdisplaybreaks\begin{align}&\lim_{\varepsilon\to0^+}\varoiint_{\partial V} f(\bm r')(\bm n'\cdot \nabla') \mathfrak g(\bm r-\varepsilon\bm n,\bm r')\D S'\notag\\={}&\lim_{\varepsilon\to0^+}\varoiint_{\partial V}f(\bm r')(\bm n'\cdot \nabla') G_D(\bm r-\varepsilon\bm n,\bm r')\D S'-\frac{1}{\varoiint_{\partial V}\D S}\varoiint_{\partial V}  f(\bm r')\D S'\notag\\{}&-2\lim_{\varepsilon\to0^+}\varoiint_{\partial V}  f(\bm r')(\bm n'\cdot \nabla')\frac{1}{4\pi|\bm r-\varepsilon\bm n-\bm r'|}\D S'\notag\\={}&-f(\bm r')-\frac{1}{\varoiint_{\partial V}\D S}\varoiint_{\partial V}  f(\bm r')\D S'-2\lim_{\varepsilon\to0^+}\varoiint_{\partial V}  f(\bm r')(\bm n'\cdot \nabla')\frac{1}{4\pi|\bm r-\varepsilon\bm n-\bm r'|}\D S'\notag\\={}&-\varoiint_{\partial V}  f(\bm r')\left[(\bm n'\cdot \nabla')\frac{1}{2\pi|\bm r-\bm r'|}+\frac{1}{\varoiint_{\partial V}\D S}\right]\D S',\quad \bm r\in V,\end{align}where we have invoked the solution to the Dirichlet boundary value  problem using $G_D$, and the Neumann boundary condition for $ G_N$. To tackle \eqref{eq:2star}, we perform the following analysis:\begin{align}&\lim_{\varepsilon\to0^+}(\bm n\cdot \nabla)\varoiint_{\partial V}  \mathfrak g(\bm r-\varepsilon\bm n,\bm r')f(\bm r')\D S'\notag\\={}&\lim_{\varepsilon\to0^+}(\bm n\cdot \nabla)\varoiint_{\partial V}  \left[G_N(\bm r-\varepsilon\bm n,\bm r')-\frac{1}{2\pi|\bm r-\varepsilon\bm n-\bm r'|}\right]f(\bm r')\D S'\notag\\={}&\lim_{\varepsilon\to0^+}(\bm n\cdot \nabla)\varoiint_{\partial V}G_N(\bm r-\varepsilon\bm n,\bm r')\left[f(\bm r)-\frac{\varoiint_{\partial V}f(\bm r'')\D S''}{\varoiint_{\partial V}\D S}\right]\D S'\notag\\&+\frac{\varoiint_{\partial V}f(\bm r'')\D S''}{\varoiint_{\partial V}\D S}\lim_{\varepsilon\to0^+}(\bm n\cdot \nabla)\varoiint_{\partial V}G_N(\bm r-\varepsilon\bm n,\bm r')\D S'\notag\\{}&-2\lim_{\varepsilon\to0^+}\varoiint_{\partial V}  f(\bm r')(\bm n\cdot \nabla)\frac{1}{4\pi|\bm r-\varepsilon\bm n-\bm r'|}\D S'\notag\\={}&-\varoiint_{\partial V}  f(\bm r')\left[(\bm n\cdot \nabla)\frac{1}{2\pi|\bm r-\bm r'|}+\frac{1}{\varoiint_{\partial V}\D S}\right]\D S',\quad\bm r\in \partial V,\end{align}where we have relied on the homogeneous Dirichlet boundary condition for $G_D$,  the solution to the Neumann boundary value problem in terms of  $G_N$, and the fact that\begin{align}&\lim_{\varepsilon\to0^+}(\bm n\cdot \nabla)\varoiint_{\partial V}G_N(\bm r-\varepsilon\bm n,\bm r')\D S'\notag\\={}&\lim_{\delta\to0^+}\lim_{\varepsilon\to0^+}(\bm n\cdot \nabla)\varoiint_{\partial V}G_N(\bm r-\varepsilon\bm n,\bm r'-\delta\bm n')\D S'\notag\\{}&+\lim_{\delta\to0^+}\lim_{\varepsilon\to0^+}(\bm n\cdot \nabla)\varoiint_{\partial V}[G_N(\bm r-\varepsilon\bm n,\bm r')-G_N(\bm r-\varepsilon\bm n,\bm r'-\delta\bm n')]\D S'\notag\\={}&-1+\lim_{\delta\to0^+}\lim_{\varepsilon\to0^+}(\bm n\cdot \nabla)\varoiint_{\partial V}[G_N(\bm r-\varepsilon\bm n,\bm r')-G_N(\bm r-\varepsilon\bm n,\bm r'-\delta\bm n')]\D S'=0.\end{align}Here, the last line can be justified as follows. For sufficiently small $ \delta >0$, we may define the boundary layer of thickness $ \delta$  as  $  L^-_{(\delta)}\colonequals \{\bm r\in V|\dist(\bm r,\partial V)<\delta\}$,   so that the function $ \dist(\bm r,\partial V)\colonequals \min_{\bm r'\in \partial V}|\bm r-\bm r'|$ is smooth in $\bm r\in L^-_{(\delta)}\cup\partial L^-_{(\delta)} $, satisfying $ (\bm n\cdot\nabla)\dist(\bm r,\partial V)=-1$. Thus, choosing $ \bm\nu'$ as the outward normal of the smooth domain $ L^-_{(\delta)}$, we have  {\begin{align}&\lim_{\delta\to0^+}\lim_{\varepsilon\to0^+}(\bm n\cdot \nabla)\varoiint_{\partial V}[G_N(\bm r-\varepsilon\bm n,\bm r')-G_N(\bm r-\varepsilon\bm n,\bm r'-\delta\bm n')]\D S'\notag\\={}&\lim_{\delta\to0^+}\lim_{\varepsilon\to0^+}(\bm n\cdot \nabla)\varoiint_{\partial V^-_{(\delta)}}[\dist(\bm r',\partial V)(\bm\nu'\cdot\nabla')G_N(\bm r-\varepsilon\bm n,\bm r')\notag\\{}&-G_N(\bm r-\varepsilon\bm n,\bm r')(\bm\nu'\cdot\nabla')\dist(\bm r',\partial V)]\D S'\notag\\={}&\lim_{\delta\to0^+}\lim_{\varepsilon\to0^+}(\bm n\cdot \nabla)\iiint_{ V^-_{(\delta)}}[\dist(\bm r',\partial V)\nabla'^{2}G_N(\bm r-\varepsilon\bm n,\bm r')\notag\\{}&-G_N(\bm r-\varepsilon\bm n,\bm r')\nabla'^{2}\dist(\bm r',\partial V)]\D^3 \bm r'\notag\\={}&-\lim_{\delta\to0^+}\lim_{\varepsilon\to0^+}(\bm n\cdot \nabla)\dist(\bm r-\varepsilon\bm n,\partial V)=1,\end{align}}as claimed.} \end{proof}

Now, for fixed $ \bm r'\in \partial V$, the boundary trace  $ (\bm n\cdot \nabla)\mathfrak g(\bm r,\bm r')$, as a function of $ \bm r\in \partial V$, has merely a weak singularity $ O(|\bm r-\bm r'|^{-1})$, hence $ (\bm n\cdot \nabla)\mathfrak g(\bm r,\bm r')\in H^{-1/2}_{\bm r}(\partial V;\mathbb C) $. {As a result,  in the independent variable  $ \bm r$, the harmonic vector field $ \nabla\mathfrak g(\bm r,\bm r')\in L^{2}_{\bm r}( V;\mathbb C^3)$ is square integrable, \textit{i.e.}~$ \mathfrak g(\bm r,\bm r')\in W^{1,2}_{\bm r}( V;\mathbb C)$ for all boundary points $ \bm r'\in \partial V$. Furthermore, $\sup_{\bm r'\in\partial V} \iiint_V|\nabla\mathfrak g(\bm r,\bm r')|^2\D^3\bm r$ is finite. The boundary trace mapping from the Sobolev space $W^{1,2}( V;\mathbb C)$ to  $ H^{1/2}(\partial V;\mathbb C)$ then leads to  $ \mathfrak g(\bm r,\bm r')\in H^{1/2}_{\bm r}(\partial V;\mathbb C),\bm r'\in  \partial V$. Moreover, the expression $ \iiint_V|\nabla\mathfrak g(\bm r,\bm r')|^2\D^3\bm r$ defines a subharmonic function in $\bm r'$, which may only attain its maximum at the boundary $ \bm r'\in \partial V$. Therefore, the condition    $\sup_{\bm r'\in\partial V} \iiint_V|\nabla\mathfrak g(\bm r,\bm r')|^2\D^3\bm r<+\infty$  entails the square integrability of $ \nabla\mathfrak g(\bm r,\bm r')\in L^{2}_{\bm r}( V;\mathbb C^3),\bm r'\in  V$. As a result, we always have   $ \mathfrak g(\bm r,\bm r')\in H^{1/2}_{\bm r}(\partial V;\mathbb C)$, no matter the point $\bm r'$ is at the boundary $\bm r'\in \partial V$ or in the interior $\bm r'\in V$.}
\begin{proposition}[A Hilbert--Schmidt operator polynomial]\label{prop:HS_proof}The operator $ (\hat I+2\hat{\mathscr G}-2\hat\gamma)^2(\hat{\mathscr G}-\hat\gamma)\colon \Phi(V;\mathbb C^3)\longrightarrow\Phi(V;\mathbb C^3)$ is of Hilbert--Schmidt type.\end{proposition}\begin{proof}Without loss of generality, we may pick the complete orthonormal basis set  $ \{\bm e_s|s=1,2,\dots\}\subset\Phi(V;\mathbb C^3)$ so that one of its subset   $ \{\bm f_s|s=1,2,\dots\}$ exhausts all the eigenvectors subordinate to the non-zero eigenvalues of the polynomially compact Hermitian operator $\hat{\mathscr G}-\hat \gamma:\Phi(V;\mathbb C^3)\longrightarrow\Phi(V;\mathbb C^3) $. The bound on the operator norm  $\Vert\hat{\mathscr G}-\hat\gamma\Vert\leq1 $ then naturally leads to \begin{align}\Vert(\hat I+2\hat{\mathscr G}-2\hat\gamma)^2(\hat{\mathscr G}-\hat\gamma)\Vert_2\leq\sqrt{\sum_{s=1}^\infty\Vert(\hat I+2\hat{\mathscr G}-2\hat\gamma)^2\bm f_s\Vert_{L^2(V;\mathbb C^3)}^2}.\end{align}In the formula above, each $ \bm f_s$ is the gradient of a harmonic function, so the following integral representations hold for $\bm F\in\Cl(\Span\{\bm f_s|s=1,2,\dots\} )$:
\begin{align}((\hat I+2\hat{\mathscr G}-2\hat\gamma)\bm F)(\bm r)={}&\iiint_V\nabla\nabla'\mathfrak g(\bm r,\bm r') \bm F(\bm r')\D^3\bm r'=\iiint_V\hat K_1(\bm r,\bm r') \bm F(\bm r')\D^3\bm r',\notag\\((\hat I+2\hat{\mathscr G}-2\hat\gamma)^{2}\bm F)(\bm r)={}&\iiint_V \nabla\nabla'\mathfrak G(\bm r,\bm r')\bm F(\bm r')\D^3\bm r'=\iiint_V\hat K_2(\bm r,\bm r') \bm F(\bm r')\D^3\bm r',\end{align}where \begin{align}\mathfrak G(\bm r,\bm r')\colonequals {}&\iiint_{V}\nabla''\mathfrak g(\bm r,\bm r'')\cdot\nabla''\mathfrak g(\bm r'',\bm r')\D^3\bm r''\end{align}evidently satisfies the harmonic equations $ \nabla^{2}\mathfrak G(\bm r,\bm r')=0$ and $\nabla'^{2}\mathfrak G(\bm r,\bm r')=0 $ for $ \bm r,\bm r'\in V$.

 Now, extending the action of the integral kernel $ \hat K_2(\bm r,\bm r')$ on     $ \{\bm f_s|s=1,2,\dots\}\subset\Phi(V;\mathbb C^3)$ to a complete orthonormal basis set of $L^2(V;\mathbb C^3) $, and using the Parseval identity on $ L^2(V;\mathbb C^3)$, we can show that\begin{align}\sum_{s=1}^\infty\Vert(\hat I+2\hat{\mathscr G}-2\hat\gamma)^2\bm f_s\Vert_{L^2(V;\mathbb C^3)}^{2}=\iiint_V\left\{\iiint_V\Tr[\hat K_2^{*}(\bm r,\bm r')\hat K_2^{\phantom{*}}(\bm r,\bm r') ]\D^3\bm r'\right\}\D^3\bm r,\label{eq:HS_bd}\end{align} where  ``$ \Tr$'' denotes the trace of a $3\times3$ matrix. To justify the equality in the formula above, we note that $ \iiint_V\hat K_2(\bm r,\bm r') \bm F(\bm r')\D^3\bm r'$  represents the gradient of a harmonic function for whatever square-integrable input $ \bm F\in L^2(V;\mathbb C^3)$, so every eigenvector $ \bm F_\lambda\in L^2(V;\mathbb C^3)$ satisfying $ \iiint_V\hat K_2(\bm r,\bm r') \bm F_\lambda(\bm r')\D^3\bm r'=\lambda\bm F_\lambda(\bm r),\lambda\neq0$ must belong to the function space $ \Phi(V;\mathbb C^3)$. In other words, the extension of the orthonormal basis set leaves the Hilbert--Schmidt norm intact.

We may go on to  cast the volume integral representation of $ \mathfrak G(\bm r,\bm r')$ into a surface integral (more precisely, a canonical pairing \cite[p.~206]{DautrayLionsVol3} between $ \mathfrak g(\bm r,\bm r'')\in H^{1/2}_{\bm r''}(\partial V;\mathbb C)$ and $(\bm n''\cdot\nabla'')\mathfrak g(\bm r'',\bm r')\in H^{-1/2}_{\bm r''}(\partial V;\mathbb C)  $) in two ways\begin{align}\mathfrak G(\bm r,\bm r')={}&\varoiint_{\partial V} \mathfrak g(\bm r,\bm r'')(\bm n''\cdot\nabla'')\mathfrak g(\bm r'',\bm r')\D S''\notag\\={}&\varoiint_{\partial V} \mathfrak g(\bm r'',\bm r')(\bm n''\cdot\nabla'')\mathfrak g(\bm r,\bm r'')\D S'',\quad\bm r,\bm r'\in V.\end{align}Clearly, the reciprocal symmetry   $ \mathfrak g(\bm r_1,\bm r_2)=\mathfrak g(\bm r_2,\bm r_1)$ entails the result  $ \mathfrak G(\bm r,\bm r')=\mathfrak G(\bm r',\bm r)$.
Interpreting the expression $ \mathfrak G(\bm r,\bm r'),\bm r\in V,\bm r'\in\partial V $ as a boundary trace (denoted by the limit notation ``$ \lim_{\varepsilon\to0^+}$'' as before), we may  use \eqref{eq:star} to deduce\begin{align}\mathfrak G(\bm r,\bm r')={}&\lim_{\varepsilon\to0^+}\varoiint_{\partial V} \mathfrak g(\bm r,\bm r'')(\bm n''\cdot\nabla'')\mathfrak g(\bm r'',\bm r'-\varepsilon\bm n')\D S''\notag\\={}&-\varoiint_{\partial V} \mathfrak g(\bm r,\bm r'')\left[(\bm n''\cdot\nabla'')\frac{1}{2\pi|\bm r'-\bm r''|}+\frac{1}{\varoiint_{\partial V}\D S'}\right]\D S'',\quad \bm r\in V,\bm r'\in \partial V.\end{align}
Then, employing the harmonic equations for $\mathfrak G(\bm r,\bm r') $, we may convert the double volume integral on the right-hand side of \eqref{eq:HS_bd} to a double surface integral in the following fashion:\begin{align}&\iiint_V\left\{\iiint_V\Tr[\hat K_2^{*}(\bm r,\bm r')\hat K_2(\bm r,\bm r') ]\D^3\bm r'\right\}\D^3\bm r\notag\\={}&\iiint_V\sum_{\bm u\in\{\bm e_x,\bm e_y,\bm e_z\}}\left\{\iiint_V\sum_{\bm v\in\{\bm e_x,\bm e_y,\bm e_z\}}[(\bm u\cdot \nabla)(\bm v\cdot\nabla')\mathfrak G(\bm r,\bm r')]^{2}\D^3\bm r'\right\}\D^3\bm r\notag\\={}&\iiint_V\sum_{\bm u\in\{\bm e_x,\bm e_y,\bm e_z\}}\left\{\varoiint_{\partial V}[(\bm u\cdot \nabla)\mathfrak G(\bm r,\bm r')][(\bm u\cdot \nabla)(\bm n'\cdot\nabla')\mathfrak G(\bm r,\bm r')]\D S'\right\}\D^3\bm r\notag\\={}&\varoiint_{\partial V}\left\{\varoiint_{\partial V}[(\bm n\cdot \nabla)\mathfrak G(\bm r,\bm r')][(\bm n'\cdot\nabla')\mathfrak G(\bm r,\bm r')]\D S'\right\}\D S.\label{eq:dV_dS}\end{align}Here, the integrands in the last line are understood in terms of boundary trace, expressible as a specific case of  \eqref{eq:2star}:\begin{align}&(\bm n\cdot \nabla)\mathfrak G(\bm r,\bm r')=\lim_{\varepsilon\to0^+}(\bm n\cdot \nabla)\mathfrak G(\bm r-\varepsilon\bm n,\bm r')\notag\\={}&\varoiint_{\partial V}  \left[ (\bm n\cdot \nabla)\frac{1}{2\pi|\bm r-\bm r''|} +\frac{1}{\varoiint_{\partial V}\D S}\right]\left[(\bm n''\cdot\nabla'')\frac{1}{2\pi|\bm r'-\bm r''|}+\frac{1}{\varoiint_{\partial V}\D S'}\right]\D S'',\;\;\bm r,\bm r'\in \partial V.\end{align}Judging from the surface integral above, the singular behavior of $ (\bm n\cdot \nabla)\mathfrak G(\bm r,\bm r')$ is comparable to the convolution of two $ O(|\bm r-\bm r'|^{-1})$  integral kernels on the boundary surface, which results in the short distance asymptotics $ (\bm n\cdot \nabla)\mathfrak G(\bm r,\bm r')=O(\log |\bm r-\bm r'|)$.  Likewise, by reciprocal symmetry   $ \mathfrak G(\bm r,\bm r')=\mathfrak G(\bm r',\bm r)$, we have
\begin{align}&(\bm n'\cdot \nabla')\mathfrak G(\bm r,\bm r')=(\bm n'\cdot \nabla')\mathfrak G(\bm r',\bm r)\notag\\={}&\varoiint_{\partial V}  \left[ (\bm n'\cdot \nabla')\frac{1}{2\pi|\bm r'-\bm r''|} +\frac{1}{\varoiint_{\partial V}\D S'}\right]\left[(\bm n''\cdot\nabla'')\frac{1}{2\pi|\bm r-\bm r''|}+\frac{1}{\varoiint_{\partial V}\D S}\right]\D S'',\end{align}which is again a surface integral kernel of order $ O(\log |\bm r-\bm r'|)$. As the logarithmic singularity is square integrable, the surface integral\begin{align}\varoiint_{\partial V}[(\bm n\cdot \nabla)\mathfrak G(\bm r,\bm r')][(\bm n'\cdot\nabla')\mathfrak G(\bm r,\bm r')]\D S'\end{align} is finite for every $\bm r\in\partial V $, it is then evident that the double surface integral in \eqref{eq:dV_dS} converges.

Hence, we have established the Hilbert--Schmidt bound $\Vert(\hat I+2\hat{\mathscr G}-2\hat\gamma)^2(\hat{\mathscr G}-\hat\gamma)\Vert_2<+\infty $. \end{proof}

\subsection*{Acknowledgments}

The author thanks Prof.~Xiaowei Zhuang (Harvard University) for her  questions on nanophotonics   in 2006, which inspired the current work. Part of this research formed   Chapter 4 and Appendix C in the  author's  PhD thesis \cite{ZhouThesis} completed in January 2010 under her supervision. The author dedicates this paper to her 50th birthday.

The author is indebted to two anonymous referees for their critical comments that enhance the presentation of the current work.


\end{document}